\documentclass[twocolumn,iop,twocolappendix]{emulateapj}
\usepackage{amssymb,amsmath}
\usepackage{mathrsfs}
\usepackage{times}
\usepackage{bm}
\usepackage{url}
\usepackage{color}

\newcommand{\beq}{\begin{equation}}
\newcommand{\eeq}{\end{equation}}

\newcommand{\msun}{\ensuremath{M_\odot}}
\newcommand{\tp}{\ensuremath{t_{\rm P}}}
\newcommand{\tw}{\ensuremath{t_{\rm w}}}
\newcommand{\texp}{\ensuremath{t_{0}}}

\newcommand{\mni}{\ensuremath{M_{\rm Ni}}}

\newcommand{\eexp}{\ensuremath{E_{\rm exp}}}
\newcommand{\mej}{\ensuremath{M_{\rm env}}}
\newcommand{\lpl}{\ensuremath{L_{\rm pl}}}

\begin{document}
\title{The Nickel Mass Distribution of Normal Type II Supernovae}
\shorttitle{The Nickel Mass Distribution of Normal Type II Supernovae}
\shortauthors{M\"uller, Prieto, Pejcha \& Clocchiatti}

\author{Tom\'as M\"uller$^{1,2 \ast}$, Jos\'e L. Prieto$^{1,3}$, Ond\v{r}ej Pejcha$^4$ and Alejandro Clocchiatti$^{1,2}$}
\affil{$^1$ Millennium Institute of Astrophysics, Santiago, Chile \\
$^2$ Instituto de Astrof\'isica, Pontificia Universidad Cat\'olica de Chile, Av. Vicuña Mackenna 4860, 782-0436 Macul, Santiago, Chile \\
		$^3$ N\'ucleo de Astronom\'ia de la Facultad de Ingenier\'ia y Ciencias, Universidad Diego Portales, Av. Ej\'ercito 441, Santiago, Chile \\        
        $^4$ Lyman Spitzer Jr. Fellow, Department of Astrophysical Sciences, Princeton University, 4 Ivy Lane, Princeton, NJ 08540, USA}
       
\email{$^{\ast}$ tmuller@astro.puc.cl}

\begin{abstract}
Core-collapse supernova explosions expose the structure and environment of massive stars at the moment of their death. We use the global fitting technique of \citet{PP15a,PP15b} to estimate a set of physical parameters of 19 normal Type II SNe, such as their distance moduli, reddenings, $^{56}$Ni masses $\mni$, and explosion energies $\eexp$ from multicolor light curves and photospheric velocity curves. We confirm and characterize known correlations between $\mni$ and bolometric luminosity at 50 days after the explosion, and between $\mni$ and $\eexp$. We pay special attention to the observed distribution of $\mni$ coming from a joint sample of 38 Type~II SNe, which can be described as a skewed-Gaussian-like distribution between $0.005~M_{\odot}$ and $0.280~M_{\odot}$, with a median of $0.031~M_{\odot}$, mean of $0.046\,\msun$, standard deviation of $0.048\,\msun$ and skewness of $3.050$. We use two-sample Kolmogorov-Smirnov test and two-sample Anderson-Darling test to compare the observed distribution of $\mni$ to results from theoretical hydrodynamical codes of core-collapse explosions with the neutrino mechanism presented in the literature. Our results show that the theoretical distributions obtained from the codes tested in this work, KEPLER and Prometheus Hot Bubble, are compatible with the observations irrespective of different pre-supernova calibrations and different maximum mass of the progenitors.
\end{abstract}


\section{Introduction}
Most massive stars with initial mass $M \gtrsim 8~M_{\odot}$ finish their lives  with a collapse of their iron cores \citep[e.g.][but see also \citealt{zapartas17} for the contribution of lower-mass stars in binary stars]{kalirai08,Smartt09,smartt+09,IH13}. A small fraction of the $\sim 10^{53}$~ergs of gravitational potential energy released in the collapse can power a core-collapse supernova (CCSN) explosion, leaving behind a neutron star or a black hole. A non-negligible fraction of massive stars might fail to explode as CCSN and instead relatively quietly collapse to a black hole (e.g., \citealt{nade80}; \citealt{burrows86}; \citealt{lieb01}; \citealt{heger03}; \citealt{kocha08}; \citealt{oconn11}; \citealt{LW13}; \citealt{kochanek14}; \citealt{adams16}; although see also \citealt{KK15} for an alternative explosion model). 

The most common kind of CCSNe are Type~II supernovae (SN~II) with broad spectral lines of hydrogen and plateau (SN~II-P) light curves (e.g., \citealt{smith11,graur17}). The success of amateur and professional supernova surveys (e.g., Cal\'an/Tololo, \citealt{hamuy93}; LOSS, \citealt{Li11}; CHASE, \citealt{pignata09}; PTF/iPTF, \citealt{rau09}; Pan-Starrs, \citealt{kaiser12}; ASAS-SN, \citealt{shapee14}) has been paramount for follow-up studies that have uncovered the full range of observed and physical properties of normal SN~II as well as significant correlations between some of their properties \citep[e.g.][]{hamuy03,arcavi12,and14,faran14,gutierrez14,sanders15,PP15a,PP15b,holoien16,valenti16,rubin16}. Hydrodynamical models of explosions of hydrogen-rich massive stars explain relatively well most of the main features of the light curves and spectra of normal SN~II \citep[e.g.][]{kasen09,bersten11,dessart11,pumo11,morozova16,lisakov17}.

Some of the CCSNe discoveries in nearby galaxies and the availability of deep pre-explosion images from HST and ground based 8-meter class telescopes have led to the detection of a number of massive star progenitors, most of them red super giants \citep[RSG; e.g.,][]{Smartt09,Smartt15}. A confrontation of these detections and upper-limits with the expectations from a normal Salpeter stellar initial mass function (IMF; \citealt{salpeter55}) constrains the main sequence progenitor masses of normal SN~II to be $8 \lesssim M
\lesssim 18~M_{\odot}$ \citep[e.g.,][]{Smartt15}. The relatively low upper limit in progenitor masses, compared to the local samples of RSG \citep[e.g.,][]{neugent12,massey16}, can be interpreted as evidence for failed explosions and black-hole formation above this mass. However, there remain other possible explanations and we need to seek a consistent picture encompassing not only the still limited set of progenitor detections, but also other constraints.

A substantial effort has been undertaken to understand the CCSN explosion mechanism with numerical simulations \citep[e.g.][and references therein]{janka12,burrows13,bruenn13,couch13,ott16}, but the ultimate goal has not been reached yet in part due to many complexities of the physics involved \citep[e.g.][]{janka16,burrows16}. As a result, the community has been developing parameterized 1D explosion models that capture some of the most important aspects of the neutrino mechanism physics. Application of these models to a wide range of progenitors has revealed that successful and failed explosions depend critically on the internal structure of the progenitors \citep[e.g.][]{oconn11,ugliano12,PT15,suk16}, producing a more complicated picture than the traditional single progenitor mass cut for failed explosions and black-hole formation \citep[e.g.][]{heger03}. These studies have also predicted the distributions of physical parameters of the supernova explosions, such as the asymptotic kinetic energies and masses of $^{56}$Ni synthesized in the explosions, which can lead to observational tests of the massive star progenitors and the explosion mechanism with complete samples of CCSNe. 

In this paper, we study the physical parameters of a sample of well-observed, normal SN~II, following the analysis by \citet{PP15a,PP15b}. We mainly focus on the observed $^{56}$Ni mass distribution and compare it with recent results from supernova explosion models. In Section~\ref{sec:dat}, we present the data of the SN~II used in this work. In Section~\ref{sec:mod}, we briefly discuss the code used to fit the multicolor light curves and expansion velocity curves. In Section~\ref{sec:fit_res}, we show the fits obtained from the code and the physical parameters. In Section~\ref{sec:analysis}, we discuss the completeness of our joint sample and focus on the nickel mass distribution. In Section~\ref{sec:conc}, we compare theoretical nickel mass distributions with our observed distribution, where we found that the KEPLER and Prometheus Hot Bubble codes seem to match the observations.


\section{Data}
\label{sec:dat}

We studied a sub-sample of 11 normal SN~II from the Calan-Tolo Supernova Program \citep[C\&T]{hamuy93} and Carnegie Type II Supernova Survey \citep[CATS]{gal16}, with sufficient photometry in the optical UBVRI bands up to the nebular phase \citep{gal16} and spectra obtained at multiple epochs in the optical wavelength range (Guti\'errez et al. 2017a,b in preparation). We obtained expansion velocities from the SNe at different epochs by measuring the position of the minimum of the P-Cygni absorption trough of the Fe~II line at rest-wavelength of 5169~\AA, which is a good tracer of the photosphere \citep{takats12}. The photometric measurements for SN~2003hn were supplemented with measurements from \citet{03hnb}.

We added 8 more well-observed, normal SN~II with data published in the literature: SN~2009ib \citep{09ib}, SN~2012ec \citep{12ec}, SN~2013ab \citep{13ab}, SN~2013ej \citep[]{13eja,13ejb},  SN~2013fs \citep[]{valenti16,Childress16,smartt+15,Yaron17}, SN~2014G \citep{14G}, ASASSN-14gm/SN~2014cx \citep[Prieto et al. 2017, in prep.]{valenti16} and ASASSN-14ha \citep{Childress16,valenti16}. Our final sample consists of 19 SN~II. The SNe with their references for the data used in this paper are presented in Table~\ref{tab:refs}.\\

\begin{deluxetable}{lccp{30mm}}
\tabletypesize{\footnotesize}
\tablecolumns{3}
\tablewidth{0pc}
\tablecaption{Supernovae and references}
\tablehead{
\colhead{Supernova} &  \colhead{Reference}  }
\startdata
SN 1992ba 	 & \citet{gal16};Guti\'errez et al. (2017a,b in prep.)\\
SN 2002gw	 & \citet{gal16};Guti\'errez et al. (2017a,b in prep.)\\
SN 2003B	 	 & \citet{gal16};Guti\'errez et al. (2017a,b in prep.)\\
SN 2003bn	 & \citet{gal16};Guti\'errez et al. (2017a,b in prep.)\\
SN 2003E	 	 & \citet{gal16};Guti\'errez et al. (2017a,b in prep.)\\
SN 2003ef	 & \citet{gal16};Guti\'errez et al. (2017a,b in prep.)\\
SN 2003fb 	 & \citet{gal16};Guti\'errez et al. (2017a,b in prep.)\\
SN 2003hd 	 & \citet{gal16};Guti\'errez et al. (2017a,b in prep.)\\
SN 2003hn	 & \citet{gal16};Guti\'errez et al. (2017a,b in prep.)\\
SN 2003ho	 & \citet{gal16};Guti\'errez et al. (2017a,b in prep.)\\
SN 2003T	 	 & \citet{gal16};Guti\'errez et al. (2017a,b in prep.)\\
SN 2009ib 	 & \citet{09ib}\\
SN 2012ec  	 & \citet{12ec}\\
SN 2013ab 	 & \citet{13ab}\\
SN 2013ej  	 & \citet{13eja};\citet{13ejb}\\
SN 2013fs	 & \citet{Childress16};\citet{smartt+15};\\
			 & \citet{valenti16};\citet{Yaron17}\\
SN 2014G 	 & \citet{14G}\\
ASASSN-14gm & Prieto et al. (2017, in prep.);\citet{valenti16}\\
ASASSN-14ha & \citet{Childress16};\citet{valenti16}

\enddata
\label{tab:refs}
\end{deluxetable}

\section{Model}
\label{sec:mod}
In order to analyze the data and derive physical parameters for the SN~II, we used the model of \citet{PP15a,PP15b}, which simultaneously fits a phenomenological model to multicolor light curves in bands ranging from the near-UV to the near-IR and photospheric expansion velocities at different epochs. The model is based on the expanding photosphere method \citep[e.g.][]{kirshner74,eastman96}, a generalization of the Baade-Wesselink technique \citep{pejchakochanek12}. Global fitting of all observations removes manual procedures such as correction for reddening and construction of $L_{\rm bol}$, and replaces them with functions that are consistently and mechanically applied to all objects. The observational uncertainties can thus be propagated through these functions revealing covariances between quantities of interest.

The global part of the model is an empirical description of the evolution of the supernova spectral energy distribution (SED), which is separated in achromatic changes of photospheric radius constrained by the expansion velocities, and chromatic changes in the SED arising from the temperature evolution constrained by photometry. \citet{PP15a} showed that such a uniquely defined sequence exists by fitting 26 nearby supernovae with data in 21 photometric bands. The solution was anchored to a couple of well-studied objects. The SED evolutionary sequence allows to utilize information from all objects in the sample, for example, by ``predicting'' the light curve in a band without actual observations and using this prediction to estimate $L_{\rm bol}$.

Each supernova is described by up to 12 parameters such as the distance modulus $\mu$ of the host galaxy, the total color excess parameterized with $E(B-V)$, the time of the explosion, the plateau duration, the transition width, and parameters specifying the phenomenological description of light curves and expansion velocities, in particular the starting point and the pace through the SED evolutionary sequence. \citet{PP15a} argued that these parameters are sufficient to describe the observed diversity of Type II-P light curves. 




The model fitting parameters are manipulated to provide the bolometric luminosity $L_{\rm bol}$ at each epoch after explosion, the ejected $^{56}$Ni mass $\mni$ from the luminosity in the radioactive decay tail, and the explosion energy $\eexp$, mass of the ejected hydrogen envelope $\mej$, and progenitor radius $R$ based on analytic scaling relations from \citet{litvinova85} and \citet{popov93} (see \citealt{PP15b} for a summary of these scaling relations). The bolometric luminosity is obtained from integrating the spectral energy distribution of the model in a given epoch between 0.19~$\mu$m (W2 band in {\em Swift/UVOT}) and 2.2~$\mu$m ($K$ band), including an extrapolation to longer wavelengths using a Rayleigh-Jeans tail of the blackbody. To constrain $\mni$ from the radioactive decay tail we assume full gamma-ray trapping.

In this work, we use the global parameters and the covariance matrix of the model constrained in \citet{PP15a}. The other input parameters for the code to fit each supernova are the magnitudes in different filters, the photospheric expansion velocities, and the explosion epochs $\texp$ with their respective uncertainties used as constraints for the fits (in some cases). These were taken from Table 4 of \citet{and14} for 4 out of 6 SN~II in common with their sample (we left $t_0$ as a free parameter for SN~2002gw and SN~2003E to get better fits), using the values with smaller uncertainties between spectral matching and explosion non-detection. For the last 8 SNe in Table~\ref{tab:refs}, $\texp$ was taken from their respective references. In some cases $\texp$ was fixed in order to constrain the fits. For some SNe, the transition time $\tw$ and one of the parameters describing the slope of the radioactive exponential decay, $\gamma_{0}$, were also fixed in order to constrain the fits (see the note in Table~\ref{tab:params}). 

\section{Results}
\label{sec:fit_res}

\subsection{Fitting results}

\begin{figure*}
\epsscale{0.9}
\plotone{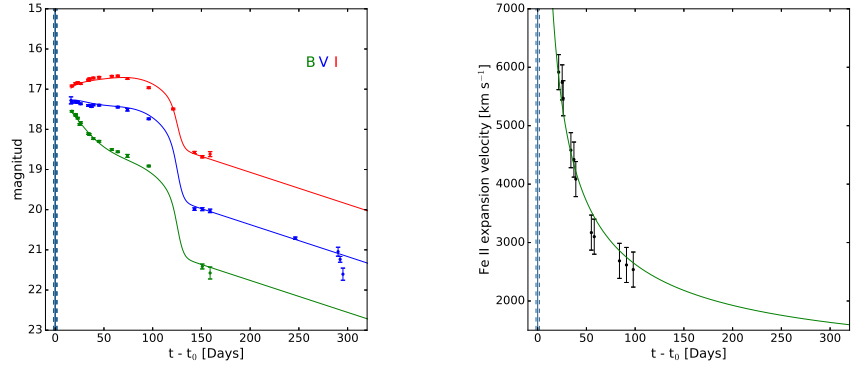}
\caption{Multiband light curves and photospheric expansion velocity curves along with their best-fit models for SN~2002gw. The left panel shows the light curves in the optical bands while the right panel shows the expansion velocity curve. The vertical solid blue line represents the explosion time $t_0$ derived from the fits with its uncertainties (vertical dashed blue lines). The complete sample is shown in Appendix.}
\label{fig:SN02gw_curves}
\end{figure*}

In Figure~\ref{fig:SN02gw_curves} we show the resulting fits to the light curves and expansion velocity curve for SN~2002gw. The entire sample of SN~II with their respective fits is shown in the Appendix. Fitting parameters for individual SN~II are given in Tables~\ref{tab:params} and~\ref{tab:params2}. We show explosion times ($t_0$ in days), plateau durations ($\tp$ in days), plateau transition widths ($\tw$ in days), total reddenings ($E(B-V)$ in mag), ejected nickel masses ($\mni$ in $M_{\odot}$), bolometric luminosities at 50 days after the explosion ($L_{\rm pl}$ in $L_{\odot}$), distance moduli ($\mu$ in mag), expansion velocity power law exponents ($\omega_1$), $\chi^2$ and the number of data points used for each fit ($n$). The code has 12 free parameters so the total number of degrees of freedom is $DOF=n-12$. The final fits and uncertainties in the parameters are obtained after renormalizing the magnitude and photospheric velocity errors to have $\chi^2/DOF=1$. This is only a crude fix assuming that the underlying model is an accurate description of the observations. 

We expected the model to reproduce well the key features of these SN~II due to previous results obtained in \citet{PP15a,PP15b}, and this is indeed what we see in the results from our work also. There are a couple of interesting cases that are worth noting. SN~2014G is the fastest declining SN~II in our sample \citep[see][for possible explanations of fast-declining SN~II]{BB92,Moriya16,moro16}. The code fits relatively well the optical light curves of this SN, except in the late-time phase of the light curve where few measurements indicate that the observed decay slope is faster than the model. This is because we are assuming full gamma-ray trapping and SN~2014G shows evidence of non-negligible leakage of gamma-rays \citep{14G}. However, the $\mni= 0.035~M_{\odot}$ estimate we obtain from our fit with full trapping is fairly consistent with the $\mni$ estimate of 0.045~M$_{\odot}$ obtained by \citet{14G}, which considers gamma-ray leakage.

Another interesting case is SN~2013fs, which was caught very close to the time of explosion and the early spectra show clear signs of strong interaction between the SN ejecta and a dense circumstellar medium or CSM \citep[e.g.,][]{Yaron17,dessart17}. In the first $\sim 10-20$~days after explosion, the model fits underestimate the SN fluxes in some bands (see lower panel of Figure~9 in the Appendix). This is likely related to extra flux produced by the strong ejecta-CSM interaction observed early on in the evolution of the SN \citep[e.g.,][]{dessart17}. The poor first of the early light curve indicate that the observed evolution was not compatible with the universal Type II-P SED evolutionary sequence described above. This potentially offers a way to diagnose similar events in the future datasets.

\begin{deluxetable*}{lllcccccc}
\tabletypesize{\footnotesize}
\tablecolumns{9}
\tablewidth{0pc}
\tablecaption{Table of fits parameters}
\tablehead{\colhead{Supernova} & \colhead{$\texp$} & \colhead{$\tp$ [d]} & \colhead{$\tw$ [d]} & \colhead{$E(B-V)$} &  \colhead{log($\mni/M_{\odot}$)} & \colhead{log($L_{\rm pl}/L_{\odot}$)}}
\startdata
SN~1992ba    & $ 48884.2	\pm 3.5 $ & $ 130.3 \pm	3.5 $ & $ 4.0 \pm	0.7 $ & $  0.175 \pm	0.013 $ &  $-1.43 \pm	0.17 $&    $8.39 \pm	0.17 $ \\
SN~2002gw   & $ 52551.6 \pm  3.2 $ & $ 125.6 \pm	6.4 $ & $  3.4 \pm	4.4 $ & $  0.113 \pm	0.018 $ &  $ -1.36 \pm	0.12 $&    $8.43 \pm	0.11 $\\
SN~2003B    & $ 52610.5 \pm  7.7 $ & $ 108.2 \pm	13.3 $ & $ 1.4 \pm	6.0 $ & $  0.025 \pm	0.019 $ &   $-1.64 \pm	0.24 $&    $8.26 \pm	0.23 $\\
SN~2003bn   & $ 52695.6 \pm  1.9 $ & $  119.2 \pm	2.1 $ & $  4.9 \pm	0.7 $ & $  0.114 \pm	0.012 $ &   $-1.51 \pm	0.09 $&    $8.36 \pm	0.08 $\\
SN~2003E    & $ 52628.0 \pm  3.4 $ & $  123.6 \pm	5.4 $ & $  2.2 \pm	1.3 $ & $  0.248 \pm	0.024 $ &   $-1.08 \pm	0.36 $&    $8.44 \pm	0.11 $\\
SN~2003ef   & $ 52743.7 \pm  5.5 $ & $ 121.9 \pm	6.1 $ & $  1.0 \pm	3.3 $ & $  0.360 \pm	0.013 $ &  $-1.04 \pm	0.14 $&   $8.67 \pm	0.11 $\\
SN~2003fb   & $ 52777.3 \pm	5.2 $ & $ 94.4 \pm	5.1 $ & $  3.6 \pm	0.6 $ & $  0.584 \pm	0.016 $ &  $-1.31 \pm	0.16 $&    $8.38 \pm	0.14 $ \\
SN~2003hd   & $ 52854.7 \pm	2.3  $ & $ 100.9 \pm	 3.0 $ & $  2.7 \pm	1.7 $ & $  0.153 \pm	0.013 $ &   $-1.52 \pm	0.08 $&    $8.44 \pm	0.08 $ \\
SN~2003hn   & $ 52870.2 \pm	1.1  $ & $ 90.5 \pm	1.1 $ & $  3.7 \pm	0.2 $ & $  0.273 \pm	0.006 $ &  $-1.74 \pm	0.09 $&   $8.28 \pm	0.09 $ \\
SN~2003ho   & $ \equiv 52847.0^{\ast}  $ & $ 80.2 \pm	1.3 $ & $ 4.8 \pm	4.6 $ & $  0.753 \pm	0.016 $ &  $-1.88 \pm	0.10 $&   $8.10 \pm	0.07 $\\
SN~2003T   & $ \equiv 52655.9^{\ast} $ & $ 104.1 \pm	0.9 $ & $  3.1 \pm	0.4 $ & $  0.239 \pm	0.013 $ &  $-1.53 \pm	0.08 $&   $8.32 \pm	0.06 $ \\
SN~2009ib   & $ \equiv 55039.0^{\ast} $ & $ \equiv 140.0^{\ast} $ & $  \equiv 5.6^{\ast} $ & $  0.179 \pm	0.006 $ &  $-1.12 \pm	0.07 $&    $8.28 \pm	0.06 $ \\
SN~2012ec   & $ \equiv 56142	.9^{\ast} $ & $  103.4 \pm 2.2 $ & $ \equiv 14.5^{\ast} $ & $  0.093 \pm	0.009 $ &   $-1.54 \pm	0.05 $&    $8.44 \pm	0.04 $\\
SN~2013ab   & $ 56342.4 \pm	1.1 $ & $ 97.2 \pm	1.0 $ & $ 4.7 \pm	0.5 $ & $  0.600 \pm	0.019 $ &  $-1.20 \pm	0.13 $&    $8.67 \pm	0.14 $ \\
SN~2013ej   & $ \equiv 56498.0^{\ast} $ & $ 100.7 \pm	0.3 $ & $  1.9 \pm	0.2 $ & $  0.163 \pm	0.009 $ &   $-2.06 \pm	0.09 $&    $8.29 \pm	0.08$\\
SN~2013fs   & $ \equiv 56571.2^{\ast} $ & $ 75.5 \pm	0.8 $ & $  7.7 \pm	0.6 $ & $  0.134 \pm	0.010 $ &   $-1.07 \pm	0.08 $&    $8.58 \pm	0.06$\\
SN~2014G   & $ \equiv 56671.3^{\ast} $ & $ 48.8 \pm	2.2 $ & $  19.6 \pm	0.7 $ & $  0.208 \pm	0.021 $ &   $-1.46 \pm	0.06 $&    $8.64 \pm	0.07$\\
ASSASSN-14gm   & $ 56901.4 \pm 0.4 $ & $ 106.7 \pm	1.0 $ & $  6.6 \pm	0.6 $ & $  0.018 \pm	0.010 $ &   $-1.12 \pm	0.09 $&    $8.56 \pm	0.08$\\
ASSASSN-14ha   & $ \equiv 56906.8 \pm 0.1 $ & $\equiv 140.0^{\ast} $ & $  1.3 \pm	0.1 $ & $  0.006 \pm	0.007 $ &   $-2.09 \pm	0.11 $&    $8.10 \pm	0.11$\\

\enddata
\tablecomments{Values with * were fixed.}
\label{tab:params}
\end{deluxetable*}

\begin{deluxetable*}{llllcccccc}
\tabletypesize{\footnotesize}
\tablecolumns{9}
\tablewidth{0pc}
\tablecaption{Table of fits parameters (\textit{continuation})}
\tablehead{\colhead{Supernova} &  \colhead{$\mu$} & \colhead{$\omega_1$} & \colhead{$\chi^2$} & \colhead{$n$}}%
\startdata
SN~1992ba   		& $ 31.36 \pm	0.39 $ & $ -0.63 \pm	0.09 $ & $ 297.3 $ & $ 52 $\\
SN~2002gw   		& $ 33.54 \pm	0.26 $ & $ -0.62 \pm	0.09 $ & $ 367.9 $ & $ 69 $\\
SN~2003B    		& $ 31.91 \pm	0.52 $ & $ -0.66 \pm	0.11 $ & $ 785.6 $ & $ 85 $\\
SN~2003bn   		& $ 33.42 \pm	0.18 $ & $ -0.82 \pm	0.07 $ & $ 198.6 $ & $ 69 $\\
SN~2003E    		& $ 34.09 \pm	0.29 $ & $ -0.84 \pm	0.07 $ & $ 286.6 $ & $ 56 $\\
SN~2003ef   		& $ 33.66 \pm	0.26 $ & $ -0.50 \pm	0.09 $ & $ 90.7 $  & $ 43 $\\
SN~2003fb   		& $ 34.05 \pm	0.36 $ & $ -0.71 \pm	0.08 $ & $ 124.1 $ & $ 47 $\\
SN~2003hd   		& $ 35.34 \pm	0.17 $ & $ -0.62 \pm	0.07 $ & $ 86.8 $  & $ 60 $\\
SN~2003hn   		& $ 30.37 \pm	0.21 $ & $ -0.75 \pm	0.01 $ & $ 2539.7$ & $ 269 $ \\
SN~2003ho   		& $ 32.35 \pm	0.17 $ & $ -0.60 \pm	0.09 $ & $ 85.7 $  & $ 40 $\\
SN~2003T    		& $ 34.69 \pm	0.16 $ & $ -0.74 \pm	0.04 $ & $ 99.0 $  & $ 50 $\\
SN~2009ib   		& $ 31.72 \pm	0.15 $ & $ -1.00 \pm	0.02 $ & $ 2420.8$ & $ 347 $ \\
SN~2012ec   		& $ 31.36 \pm	0.09 $ & $ -0.77 \pm	0.03 $ & $ 3404.8$ & $ 294 $\\
SN~2013ab   		& $ 31.33 \pm	0.31 $ & $ -0.38 \pm	0.03 $ & $ 6311.0$ & $ 287 $ \\
SN~2013ej   		& $ 28.96 \pm	0.20 $ & $ -0.68 \pm	0.01 $ & $ 8518.7$ & $ 559 $\\
SN~2013fs   		& $ 33.26 \pm	0.13 $ & $ -0.83 \pm	0.01 $ & $ 7158.7$ & $ 411 $\\
SN~2014G    		& $ 31.81 \pm	0.15 $ & $ -0.70 \pm	0.02 $ & $ 16429.4$ & $ 404 $\\
ASSASSN-14gm    & $ 31.78 \pm	0.20 $ & $ -0.70 \pm	0.03 $ & $ 4978.8$  & $ 620 $\\
ASSASSN-14ha    & $ 30.83 \pm	0.26 $ & $ -0.84 \pm	0.01 $ & $ 16480.8$ & $ 917 $

\enddata
\label{tab:params2}
\end{deluxetable*}

\subsection{Physical Parameters}

In Table~\ref{tab:deriv_params} we present the explosion energies ($E_{\rm exp}$ in ergs), ejected hydrogen envelopes masses ($\mej$ in $M_{\odot}$), and progenitors radii ($R$ in $R_{\odot}$) derived from the parametrized scaling relations in \citet{litvinova85} and \citet{popov93}. Both use a linear relation of the form: 

\begin{equation}
\text{log}(\eexp/10^{50}\text{ergs}) = \alpha \cdot \text{b} + \eta
\label{Eq:1}
\end{equation}

where b = ($M_V$,log $\tp$,log $v$), $M_V$ is the absolute magnitude in the $V$ band at 50~days, $\tp$ is the duration of the plateau phase, and $v$ is the expanding velocity of the photosphere at 50~days in units of 1000~km~s$^{-1}$. Similar relations are used for $\mej$ and $R$, but using their respective values of $\alpha$ and $\eta$ (see \citealt{litvinova85} and \citealt{popov93} for more details). It is worth noticing that the radii obtained with these scaling relations from the light curve fits are on average smaller ($300-500$~R$_{\odot}$) compared to the values estimated from observed RSG of SN~II \citep[e.g.,][]{levesque05}. Other works that have studied the radii of SN~II progenitors from their very early light curves obtain similar results \citep[e.g.,][]{GG15}. The discrepancy is understandable, because at mid-plateau, where the quantities relevant for scaling relations are obtained, the supernova photosphere has substantially retreated inward in the mass coordinate and the radius from the scaling relation thus does not represent the original surface of the star. Nonetheless, the radii of the progenitor estimated from the scaling relations do not affect the estimates of $\mni$, which estimation comes from the radiative tail.

For further analysis we combined our sample with the updated sample of \citet{PP15b}, doubling the number of objects to a total of 38. Some of the main parameters for this joint sample have the following ranges: ${\rm log}(\mni/M_{\odot})$ = $[-2.34,-0.55]$~dex, with mean $-1.52$~dex and dispersion $0.41$~dex, ${\rm log}(L_{\rm pl}/L_{\odot})$ = $[7.55,9.08]$~dex, with mean $8.38$~dex and dispersion $0.29$~dex, $\tp$ = $[48.8,140.0]$~days, with mean $110.2$~days and dispersion $20.9$~days, $\tw$ = $[0.1,27.3]$~days with mean $5.6$~days and dispersion $5.9$~days, and $\omega_1$ = $[-1.00,-0.33]$, with mean $-0.70$ and dispersion $0.16$.

\begin{figure}[h]
\epsscale{0.6}
\centering
\includegraphics[width=\columnwidth]{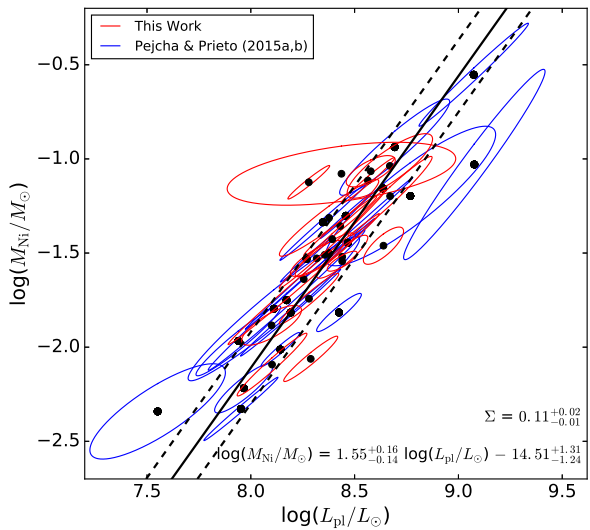}
\caption{The correlation between the plateau luminosity at 50 days after the explosion, $\lpl$, and the nickel mass, $\mni$, for the joint sample (our sample in red plus the sample of \citealt{PP15a,PP15b} in blue). We show the best linear fit and the intrinsic width of the relation with solid and dashed lines, respectively.}
\label{fig:mni}
\end{figure}

In Figure~\ref{fig:mni} we show the estimates of $\lpl$ versus $\mni$ of this joint sample, with their 1$\sigma$ confidence ellipsoids obtained from the covariances between these parameters calculated by the fitting code. We confirm the known correlation between $\lpl$ and $\mni$ \citep[e.g.][]{hamuy03,spiro14,PP15a,PP15b}. We calculated the best linear fit taking into account the covariance matrix of each data point and the intrinsic width of the relation using the generating function of \citet{hogg10} and the MCMC sampler {\tt emcee} \citep{fore13}. The best-fit relation is given in Figure~\ref{fig:mni}. Our slope is compatible within the uncertainties with the results of \citet{PP15b}. The intrinsic width of the relation is $\Sigma = 0.11^{+0.02}_{-0.01}$, which implies a scatter of $0.2$ dex in $\mni$ for fixed $\lpl$.

\begin{figure*}
\epsscale{1.}
\plottwo{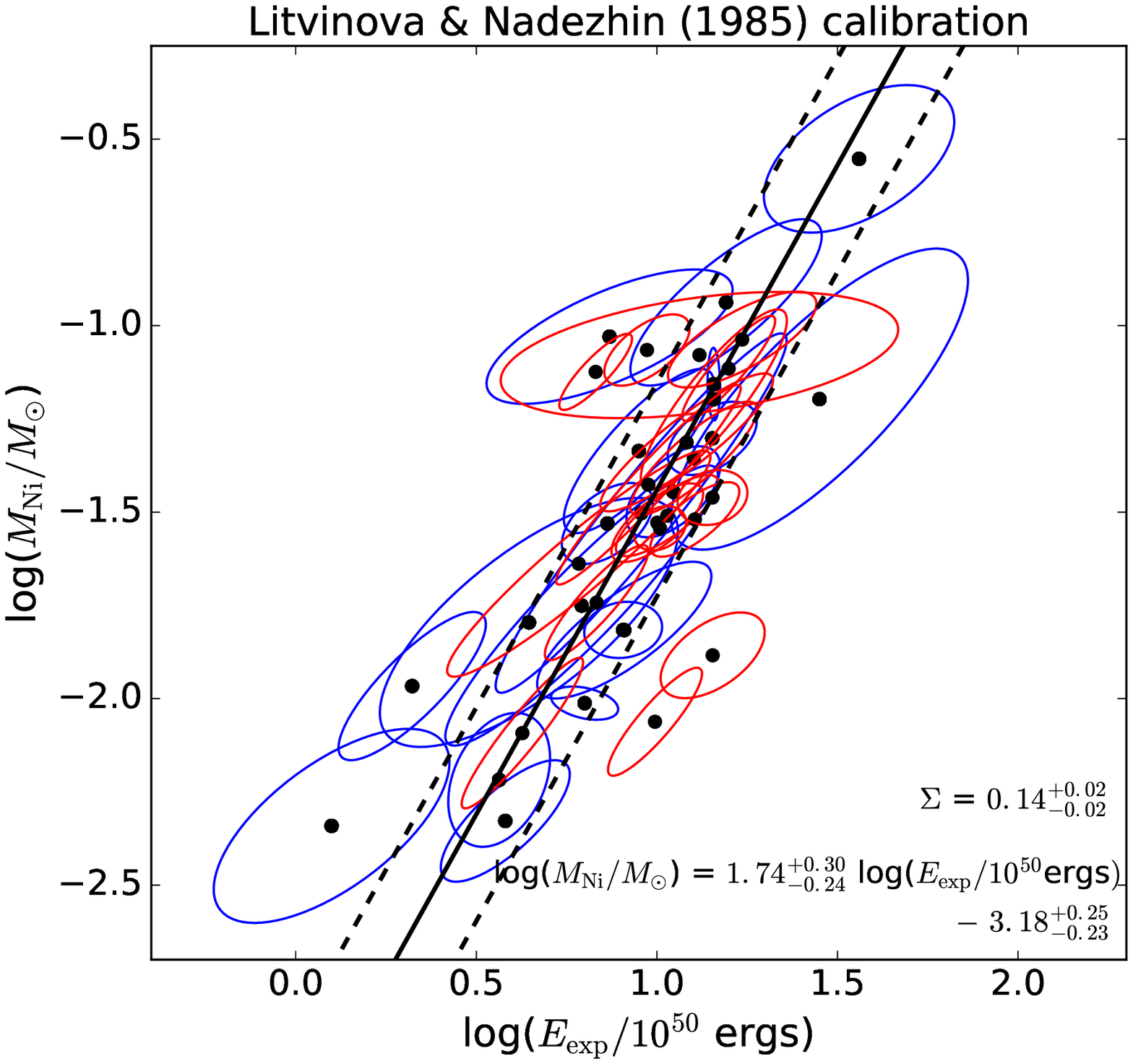}{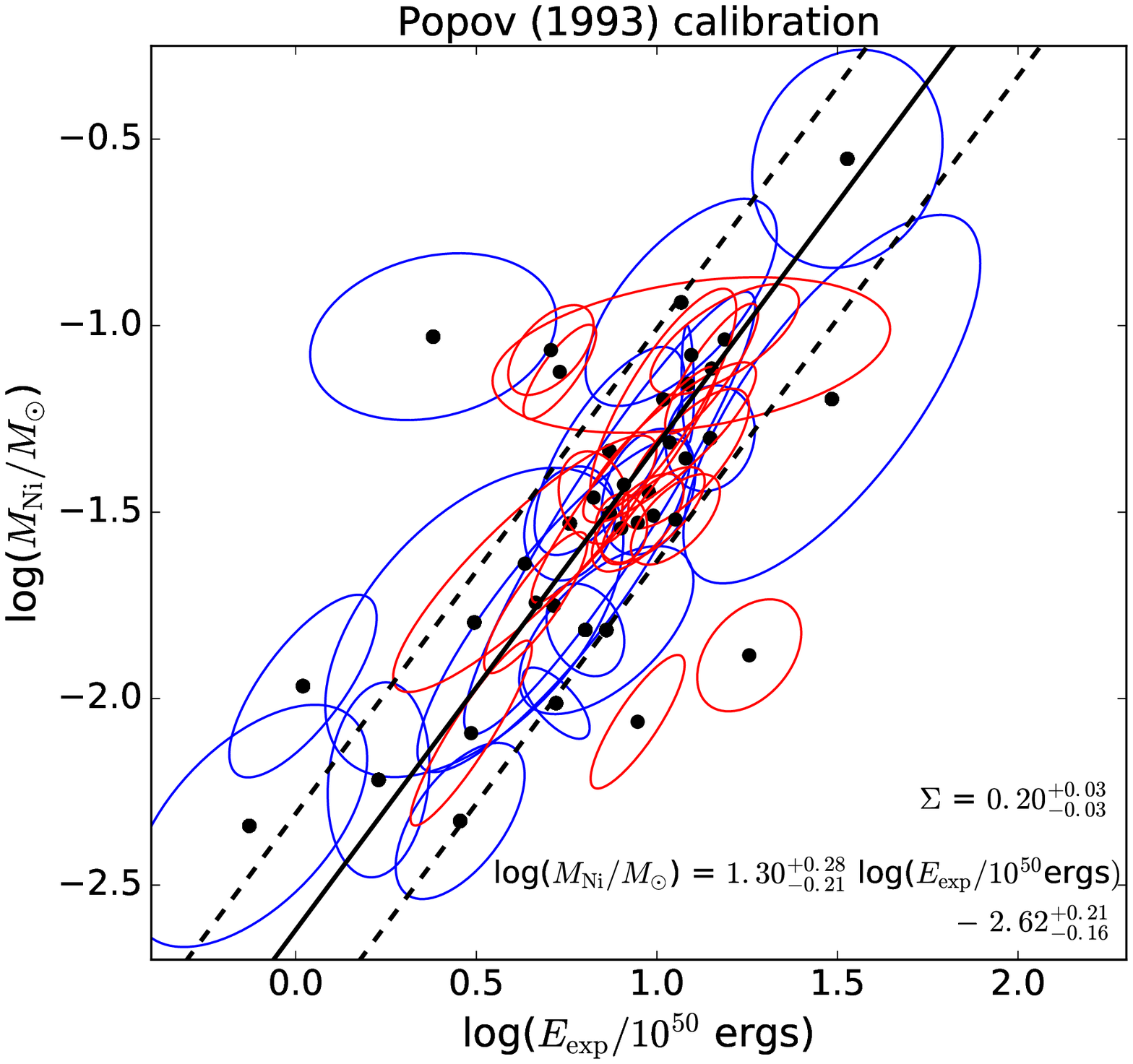}
\caption{Nickel mass, $\mni$, as a function of explosion energy, $\eexp$, for the joint sample with the scaling relations of \citet[\em{left panel}]{litvinova85} and \citet[\em{right panel}]{popov93}. We can appreciate the correlation between $\mni$ and $\eexp$, although it is not as evident as the one between $\lpl$ and $\mni$ from Fig.~\ref{fig:mni}. We show the best linear fit and the intrinsic width for both scaling relations.}
\label{fig:m_e_PP}
\end{figure*}

The relation between $\mni$ and $\eexp$ is shown in Figure~\ref{fig:m_e_PP} for the joint sample, with the 1$\sigma$ confidence ellipsoids for each SN. In the left panel we show the results for the scaling relations in \citet{litvinova85} and in the right panel for \citet{popov93}. We see that there are some differences between the two scaling relations, but the relative positions of the majority of the points are almost unchanged. The slopes and intrinsic widths are compatible within uncertainties with the results of \citet{PP15b}. The intrinsic widths are large, specially for the scaling relation of \citep{popov93}, but we do not have a reason to trust one over the other. 

We can also see in Figure~\ref{fig:m_e_PP} that the observations show larger scatter in $\eexp$ than the theoretical models (see Figure 17 of \citealt{suk16}), which could provide interesting constraints into the explosion mechanism that would be worth exploring by future models. The measured scatter might in part be caused by the scaling relations, however, a scatter of $\sim 1$ dex for $\mni \approx 0.01 \msun$ is hardly explained by uncertainties in them. In the case of the $\mej$ estimates from the scaling relations, the results could indeed be biased because the scalings do not take into account the contribution of the He core material, which in massive progenitors represents a substantial fraction of the ejecta mass. The employed scaling relations also do not include corrections to $t_{\rm P}$ from $M_{\rm Ni}$ \citep{kasen09,suk16}.
Further analysis in this direction is out of the scope of this work.

\begin{deluxetable*}{l|ccc|ccc}
\tabletypesize{\footnotesize}
\tablecolumns{7}
\tablewidth{0pc}
\tablecaption{Table of results for derived parameters}
\tablehead{\colhead{} & \multicolumn{3}{|c|}{\citet{litvinova85}} &\multicolumn{3}{c}{\citet{popov93}}   \\
{Supernova} & {$\log(E_{\rm exp}/10^{50}ergs)$} & {$\log(\mej/M_{\odot})$} & {$\log(R/R_{\odot})$} & {$\log(E_{\rm exp}/10^{50}ergs)$} & {$\log(\mej/M_{\odot})$} & {$\log(R/R_{\odot})$}
}
\startdata

SN1992ba 	 & $ 0.98 \pm 0.18 $ & $ 1.40 \pm 0.06 $ & $ 2.48 \pm 0.06 $ & $ 0.91 \pm 0.21 $ & $ 1.21 \pm 0.08 $ & $ 2.73 \pm 0.07 $  \\
SN2002gw 	 & $ 1.10 \pm 0.10 $ & $ 1.44 \pm 0.06 $ & $ 2.42 \pm 0.05 $ & $ 1.08 \pm 0.12 $ & $ 1.26 \pm 0.08 $ & $ 2.64 \pm 0.07 $  \\ 
SN2003B 	 & $ 0.78 \pm 0.20 $ & $ 1.21 \pm 0.11 $ & $ 2.43 \pm 0.11 $ & $ 0.63 \pm 0.23 $ & $ 0.96 \pm 0.14 $ & $ 2.71 \pm 0.14 $  \\ 
SN2003bn 	 & $ 1.03 \pm 0.07 $ & $ 1.38 \pm 0.03 $ & $ 2.38 \pm 0.04 $ & $ 0.99 \pm 0.09 $ & $ 1.20 \pm 0.04 $ & $ 2.60 \pm 0.06 $  \\ 
SN2003E 	 & $ 1.12 \pm 0.11 $ & $ 1.43 \pm 0.06 $ & $ 2.41 \pm 0.06 $ & $ 1.09 \pm 0.14 $ & $ 1.26 \pm 0.07 $ & $ 2.64 \pm 0.08 $  \\ 
SN2003ef 	 & $ 1.24 \pm 0.09 $ & $ 1.42 \pm 0.04 $ & $ 2.54 \pm 0.07 $ & $ 1.19 \pm 0.10 $ & $ 1.22 \pm 0.05 $ & $ 2.81 \pm 0.09 $  \\ 
SN2003fb 	 & $ 1.08 \pm 0.12 $ & $ 1.26 \pm 0.06 $ & $ 2.26 \pm 0.08 $ & $ 1.03 \pm 0.14 $ & $ 1.06 \pm 0.07 $ & $ 2.48 \pm 0.09 $  \\ 
SN2003hd 	 & $ 1.10 \pm 0.07 $ & $ 1.29 \pm 0.03 $ & $ 2.34 \pm 0.05 $ & $ 1.05 \pm 0.08 $ & $ 1.09 \pm 0.05 $ & $ 2.57 \pm 0.07 $  \\ 
SN2003hn 	 & $ 0.83 \pm 0.10 $ & $ 1.10 \pm 0.04 $ & $ 2.38 \pm 0.03 $ & $ 0.66 \pm 0.12 $ & $ 0.83 \pm 0.05 $ & $ 2.67 \pm 0.04 $  \\ 
SN2003ho 	 & $ 1.15 \pm 0.08 $ & $ 1.30 \pm 0.05 $ & $ 1.83 \pm 0.05 $ & $ 1.26 \pm 0.10 $ & $ 1.18 \pm 0.06 $ & $ 1.90 \pm 0.07 $  \\ 
SN2003T 	 & $ 1.00 \pm 0.07 $ & $ 1.30 \pm 0.03 $ & $ 2.30 \pm 0.03 $ & $ 0.95 \pm 0.09 $ & $ 1.10 \pm 0.04 $ & $ 2.52 \pm 0.04 $  \\ 
SN2009ib 	 & $ 0.83 \pm 0.07 $ & $ 1.39 \pm 0.02 $ & $ 2.52 \pm 0.03 $ & $ 0.73 \pm 0.08 $ & $ 1.20 \pm 0.03 $ & $ 2.79 \pm 0.04 $  \\ 
SN2012ec 	 & $ 1.01 \pm 0.04 $ & $ 1.25 \pm 0.04 $ & $ 2.43 \pm 0.04 $ & $ 0.90 \pm 0.06 $ & $ 1.02 \pm 0.05 $ & $ 2.70 \pm 0.06 $  \\ 
SN2013ab 	 & $ 1.16 \pm 0.16 $ & $ 1.21 \pm 0.06 $ & $ 2.56 \pm 0.06 $ & $ 1.02 \pm 0.20 $ & $ 0.94 \pm 0.07 $ & $ 2.88 \pm 0.08 $  \\ 
SN2013ej 	 & $ 1.00 \pm 0.10 $ & $ 1.28 \pm 0.03 $ & $ 2.26 \pm 0.03 $ & $ 0.95 \pm 0.12 $ & $ 1.09 \pm 0.04 $ & $ 2.47 \pm 0.04 $  \\
SN2013fs 	 & $ 0.97 \pm 0.06 $ & $ 0.95 \pm 0.03 $ & $ 2.58 \pm 0.03 $ & $ 0.71 \pm 0.08 $ & $ 0.59 \pm 0.03 $ & $ 2.96 \pm 0.04 $  \\
SN2014G 		 & $ 1.15 \pm 0.05 $ & $ 0.69 \pm 0.06 $ & $ 2.54 \pm 0.07 $ & $ 0.83 \pm 0.08 $ & $ 0.26 \pm 0.09 $ & $ 2.95 \pm 0.10 $  \\
ASASSN-14gm	 & $ 1.20 \pm 0.09 $ & $ 1.34 \pm 0.03 $ & $ 2.42 \pm 0.03 $ & $ 1.15 \pm 0.11 $ & $ 1.14 \pm 0.04 $ & $ 2.67 \pm 0.04 $  \\
ASASSN-14ha	 & $ 0.63 \pm 0.13 $ & $ 1.33 \pm 0.04 $ & $ 2.50 \pm 0.03 $ & $ 0.49 \pm 0.16 $ & $ 1.12 \pm 0.06 $ & $ 2.78 \pm 0.04 $ 
\enddata
\label{tab:deriv_params}
\end{deluxetable*}

\section{Analysis}
\label{sec:analysis}

\subsection{Sample completeness}
\label{sec:sam_com}

Our joint sample is potentially biased, because the objects come from several different surveys and we selected objects based purely on good photometric multi-wavelength coverage, including the radioactive decay phase, and availability of several epochs of optical spectroscopic data to derive photospheric expansion velocity. To assess the completeness of our joint SN~II sample, we compare it to the volume-limited sample of \citet{Li11} based on the Lick Observatory Supernova Search (LOSS). LOSS sample should be more complete, with all the discoveries coming from the same survey.

In order to compare the two samples, we used the peak absolute $R$-band magnitudes of SN~II (II-P and II-L) reported in \citet{Li11}. However, the reported peak absolute magnitudes in that study were not corrected for internal extinction in their host galaxies, so we used the average extinction from our joint sample ($A_R = 0.44$~mag) as an approximate extinction correction for the LOSS SN~II sample. Also, some of their low luminosity SN~II seemed to be caused by an above average host extinction due to their position in their host galaxies and we took them out of the sample. After these corrections, the lowest luminosity objects in \citet{Li11} sample are SN~1999br \citep{pasto04} and SN~2003Z \citep{spiro14}. In our joint sample, the lowest luminosity SN~II is SN~2001dc \citep{pasto04}. Comparing both samples, we found a slight deficit of low luminosity SN~II in our joint sample, but overall the distributions of peak absolute magnitudes are fairly consistent. The minimum, maximum, and average peak absolute magnitudes for our joint sample are: $M_{R}=-18.9$, $-14.7$ and $-16.9$~mag, respectively. For the sample of \citet{Li11}, these values are: $M_{R} = -19.1$, $-13.9$, and $-16.8$, respectively.

\subsection{$\mni$ distribution: observations vs theory}

The iron peak isotope $^{56}$Ni is synthesized in the supernova explosion within the inner few 1000~km of the progenitor, making it a sensitive probe of the explosion and other uncertain physics \citep[e.g.,][]{PT15}. Therefore, we are interested in comparing the distributions of $\mni$ from the observational (joint) sample with theoretical results from CCSNe explosion models. Our distribution of $\mni$ can be described as a skewed-Gaussian-like distribution between $0.005~M_{\odot}$ and $0.280~M_{\odot}$, with a median of $0.031~M_{\odot}$, mean of $0.046\,\msun$, standard deviation of $0.048\,\msun$, and a skewness of $3.050$.

We use the recent results by \citet{suk16} on parametrized CCSNe explosion models from the neutrino mechanism as a basis for comparison with observations. This study is particularly well-suited for comparing with the observations because they present a grid of explosion models starting from 200 progenitor masses in the range $9-120\, M_{\odot}$. The set of parameters for the different models are calibrated on the observed properties of SN~1987A for progenitors with $M > 12~M_{\odot}$ and SN~1054 (the Crab) for progenitors with $M \leq 12~M_{\odot}$. 

These progenitors were exploded with two different hydrodynamic codes, Prometheus-Hot Bubble \citep[P-HOTB;][]{JM96,kifo03} and KEPLER \citep{wea78} with their physics fully discussed in the literature \citep[e.g.,][]{woos02,WH07,SW14,WH15}. The P-HOTB code includes the neutrino and high-density physics to follow iron-core collapse, neutrino energy and lepton-number transport, while the KEPLER code does not include neutrino transport but is capable of calculating detailed nucleosynthesis and light curves.

The explosions were produced by: (1) use KEPLER to calculate the evolution of a ZAMS star until pre-SN; export results to P-HOTB, (2) use P-HOTB to calculate the collapse and neutrino-transport, provide a range of plausible $\mni$; export results back to KEPLER, and (3) use KEPLER to calculate the nucleosynthesis and light curves, adjust certain parameters to give $\mni$ roughly in the middle of the range predicted by P-HOTB. This procedure resulted in a range of possible $\mni$, within which the expected true value should be, given by P-HOTB and a single value within that range for KEPLER \citep[see section 3.2 in][for more details]{suk16}, for each progenitor. Due to the employed procedure, $\mni$ from KEPLER and P-HOTB are not independent. We took this data set from the online model database associated with this paper\footnote{\url{wwwmpa.mpa-garching.mpg.de/ccsnarchive/data/SEWBJ{\_}2015/index.html}}.

In order to construct a theoretical distribution of $\mni$, we started by assuming a Salpeter IMF with $dN/dM \propto M^{-2.35}$ for the massive star progenitors, a reasonable assumption in this mass range \citep{bastian10}. Then, we randomly selected 100,000 progenitor masses from a Salpeter IMF between $M_{\rm min}$ and $M_{\rm max}$, where $M_{\rm min}$ was kept fixed at $9~M_{\odot}$ (given by the minimum progenitor mass studied in \citealt{suk16}) and $M_{\rm max}$ was initially set at $20~M_{\odot}$ to be consistent with the constraints from SN~II progenitors \citep[e.g.,][]{Smartt15}. 

For each progenitor with a successful explosion ($\mni > 0$), for which we assigned the nearest neighbor for masses between two values of their grid, we assigned two values of $\mni$. One of the $\mni$ values was obtained from KEPLER tabulated results and the other was obtained using a random uniform distribution within the range of $\mni$ given by P-HOTB (see Figure 12 from \citealt{suk16}). We also tried a linear interpolation. In this case, for progenitor masses between two values within the grid, we assign $\mni$ values by using a linear interpolation between the $\mni$ values associated to the progenitor masses in the grid. Using the interpolation resulted in similar distributions, so we do not show this results in this work. We used the Z9.6 progenitor model calibration (Crab-like) for $M \leq 12~M_{\odot}$, together with N20 and W18 progenitor model calibrations (SN~1987A-like) for $M > 12~M_{\odot}$, because they characterize best the progenitor of SN~1987A according to \citet{suk16}. It is worth noting that the W18 model calibration produces a slightly higher fraction of failed explosions than N20 (Figure 13 of their work).

\begin{figure*}
\epsscale{1.17}
\plottwo{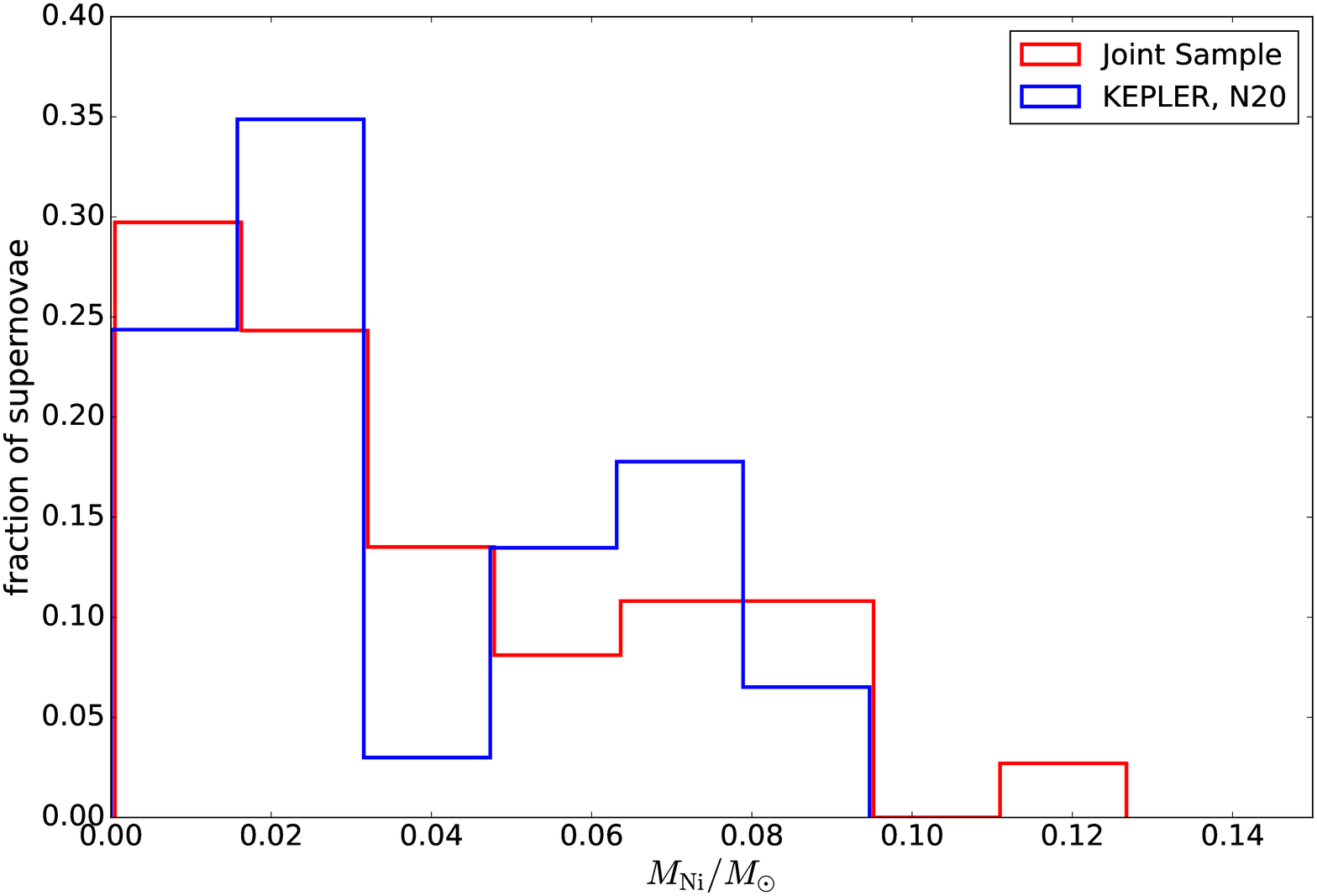}{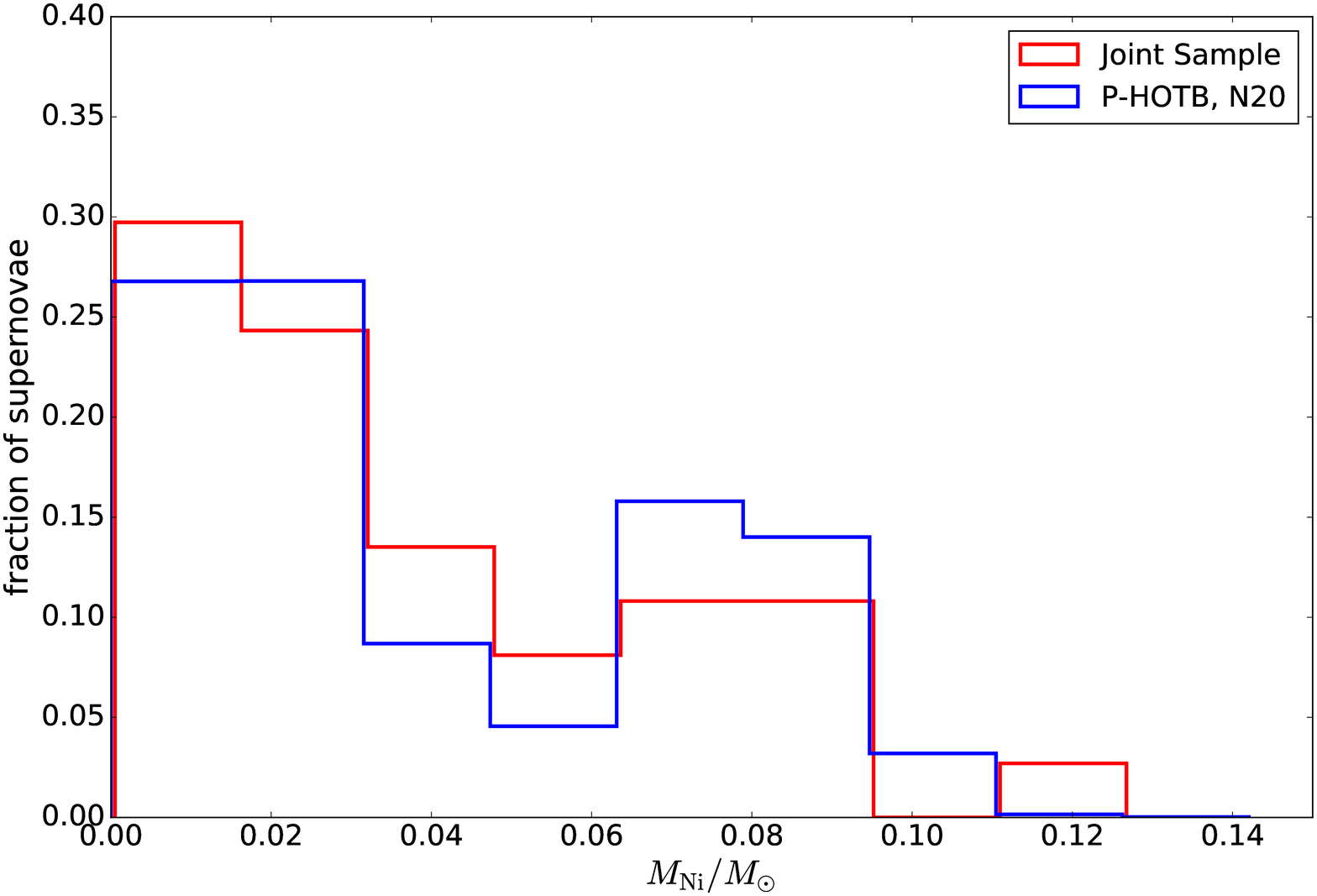}
\plottwo{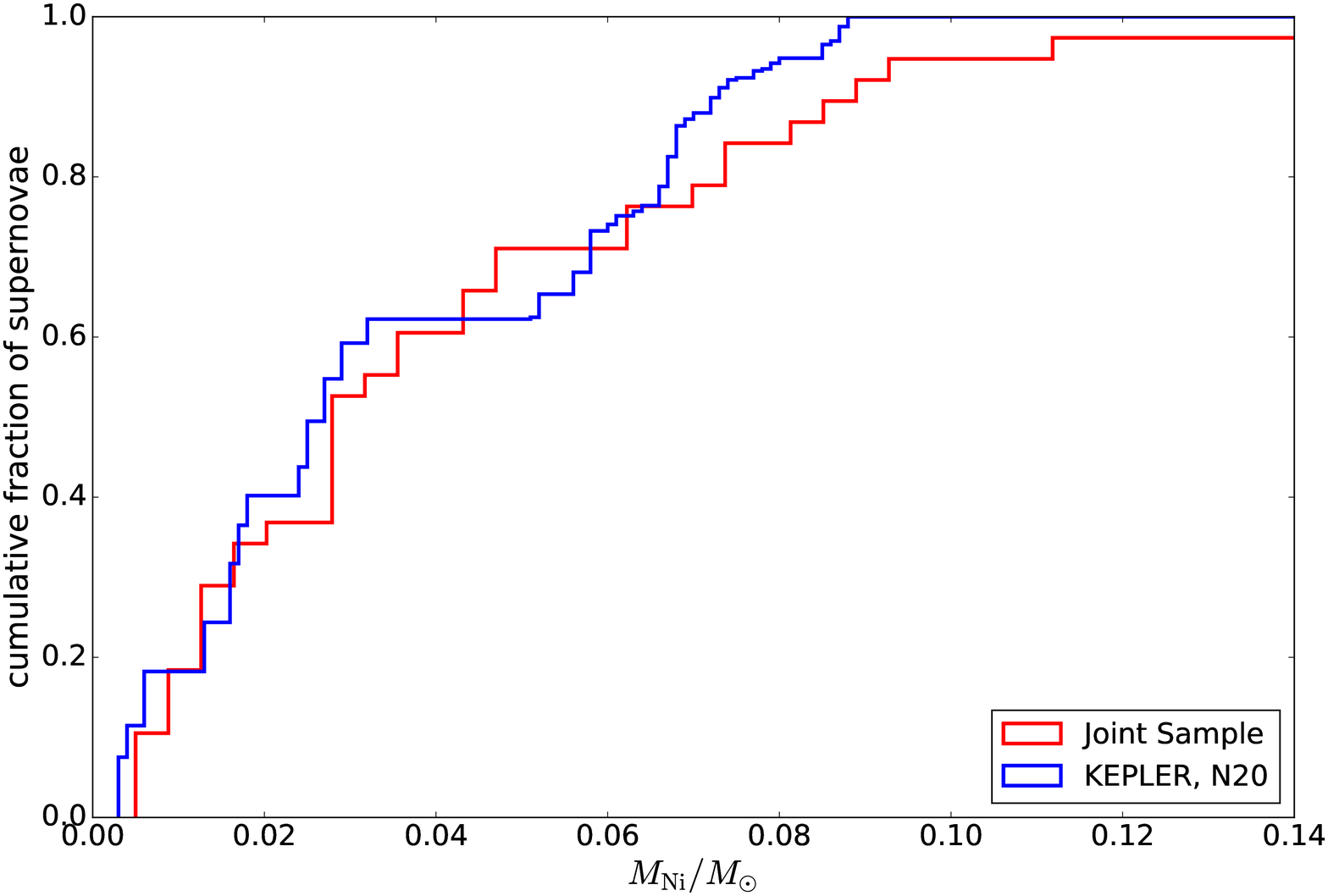}{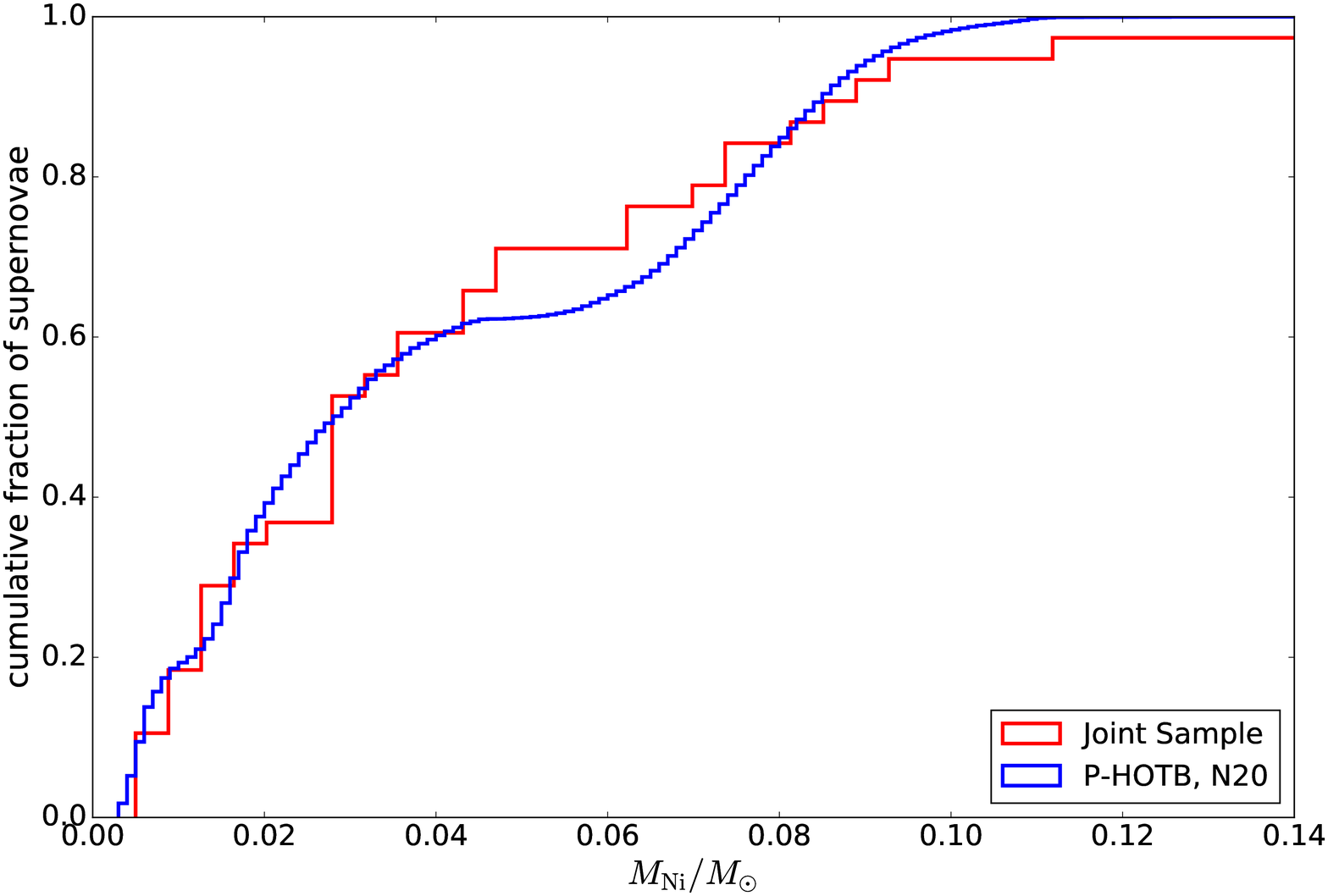}
\caption{Comparison of the observed distribution of nickel mass $\mni$ of our SN~II sample with the theoretical distribution obtained from CCSN explosion models of \citet{suk16}. The theoretical distributions were calculated for two different hydrodynamical codes: KEPLER ({\em left panels}) and P-HOTB ({\em right panels)}; for progenitors following a Salpeter IMF with $M_{\rm max} =$ 20 $M_{\odot}$ and $M_{\rm min} =$ 9 $M_{\odot}$. In the {\em upper panels}, we show the comparison of $\mni$ distributions for the KEPLER, with a small offset for visualization purposes, and P-HOTB codes with the joint sample. In the {\em lower panels}, we show the same, but for the cumulative distributions. We used the N20 calibration for progenitors with $M > 12~M_{\odot}$ and the only calibration available for progenitors with $M \leq 12~M_{\odot}$, Z9.6. The nickel yield from SN~1992H of $0.280~M_{\odot}$ is not included.}
\label{fig:histo_N20}
\end{figure*}

\begin{figure*}
\epsscale{1.17}
\plottwo{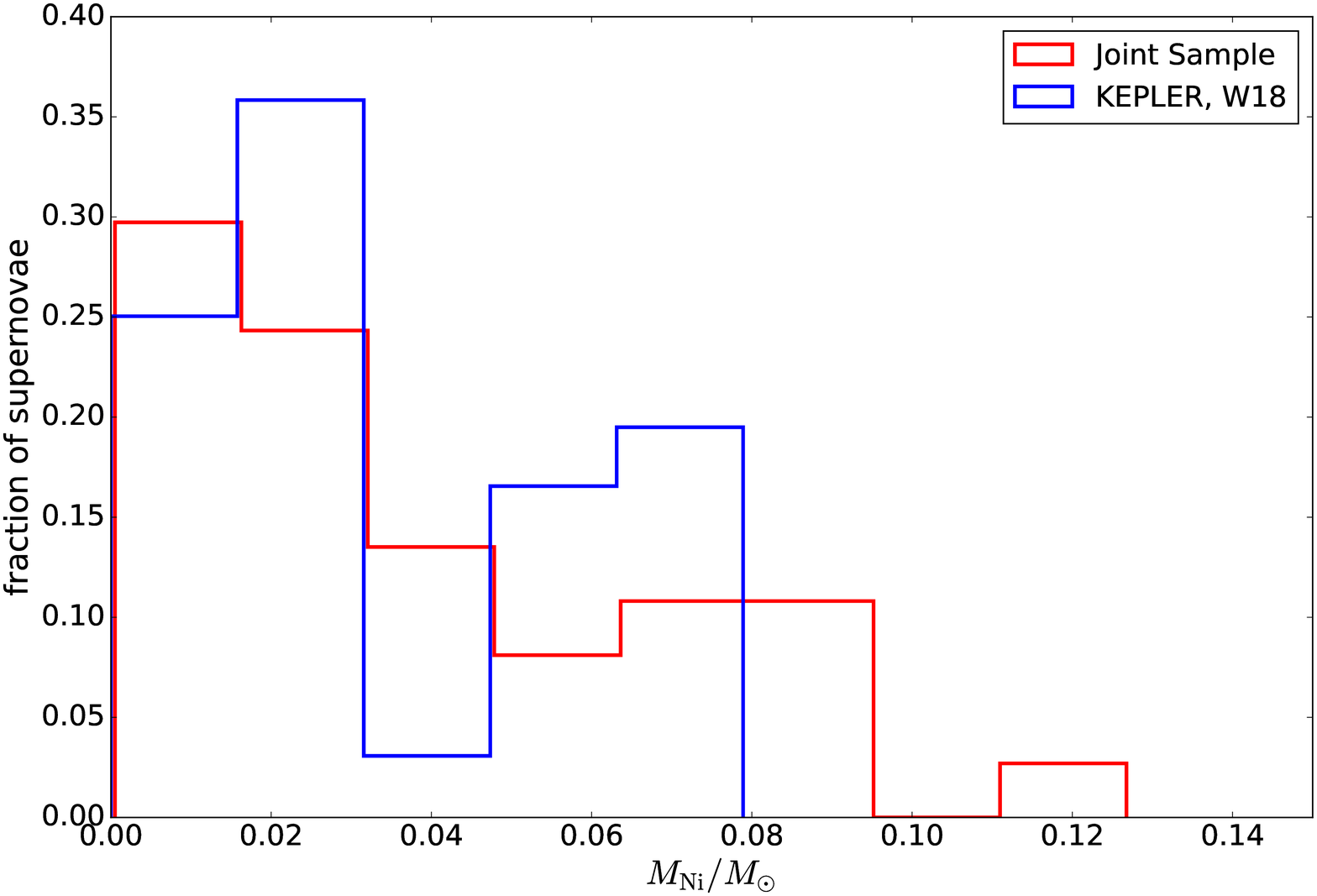}{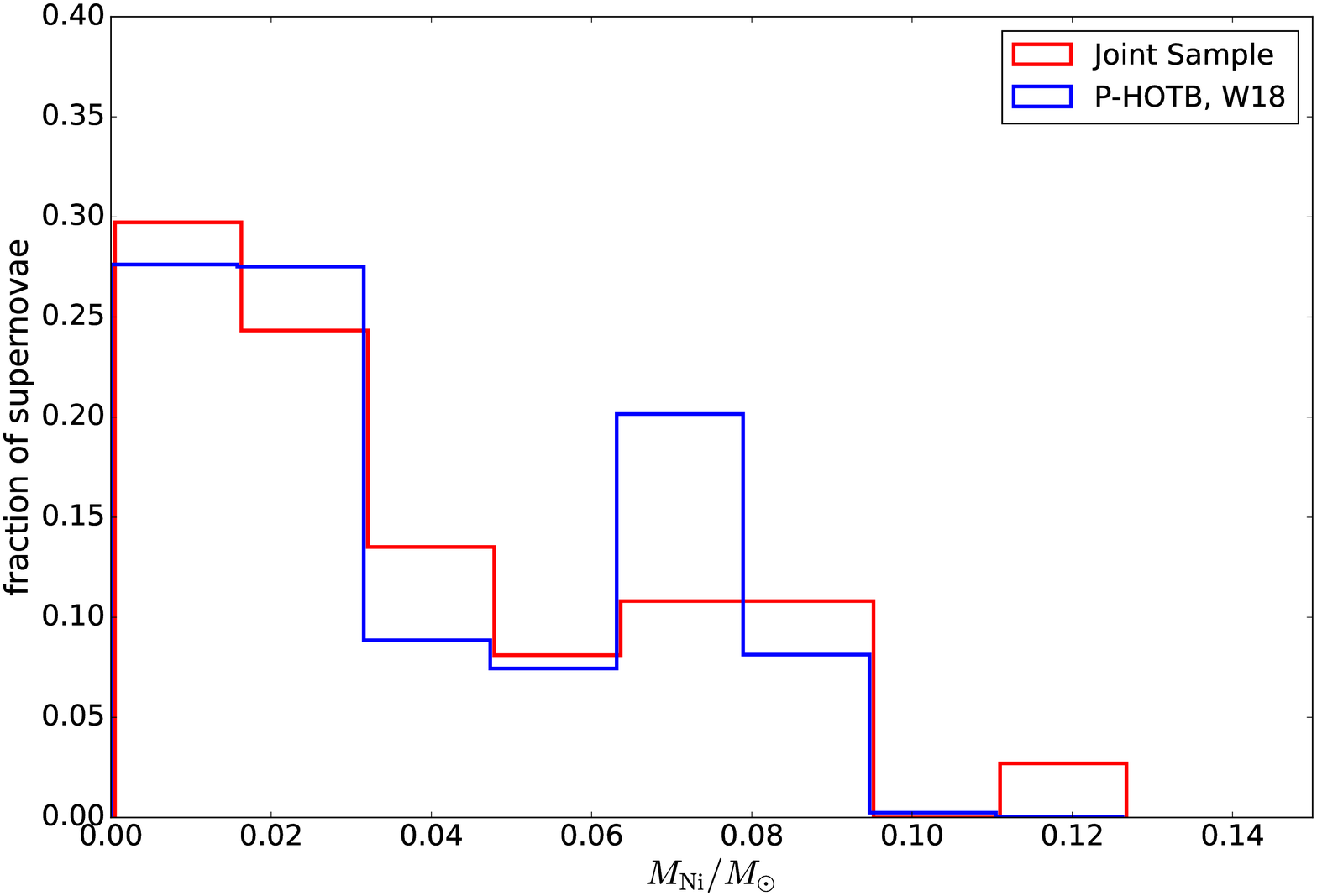}
\plottwo{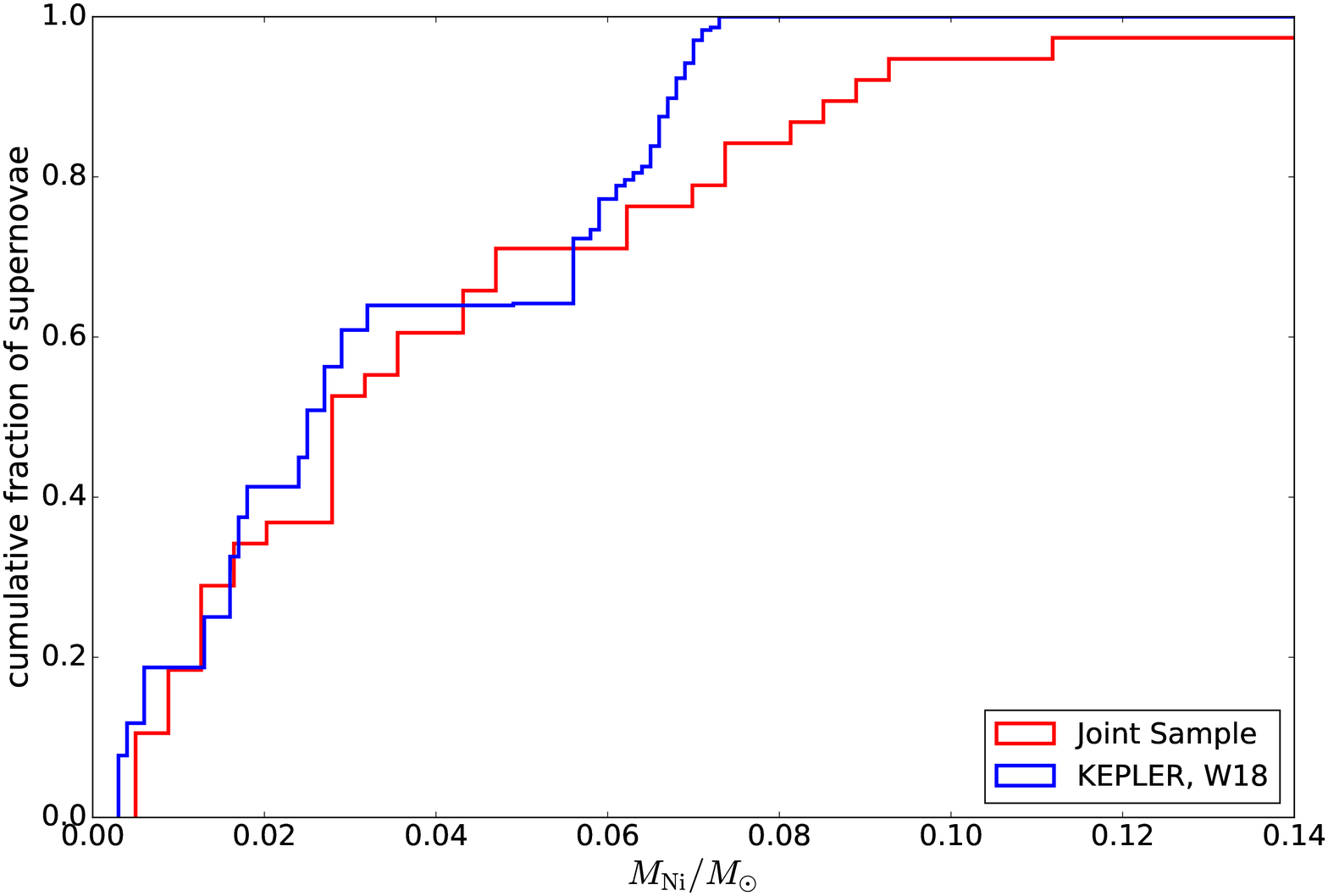}{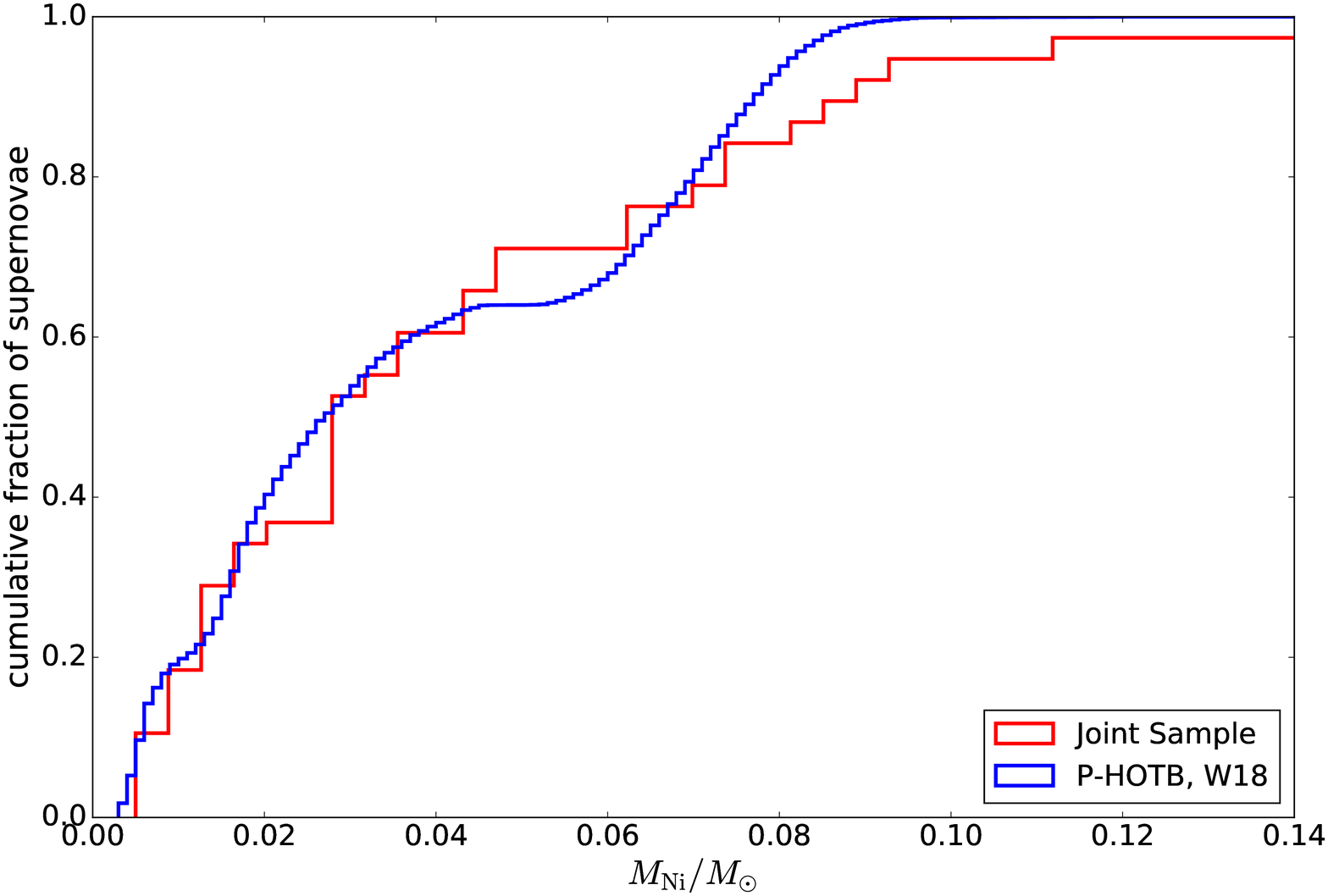}
\caption{Same as in Figure~\ref{fig:histo_N20}, but for the W18 calibration presented in \citet{suk16}. We see that in this case the theoretical distributions show lower $\mni$ values compared with the distributions obtained with the N20 calibration.}
\label{fig:histo_W18}
\end{figure*}

In the upper panels of Figure~\ref{fig:histo_N20} we show the comparison of $\mni$ distributions between the observational sample and the theoretical distributions obtained from the KEPLER (left) and P-HOTB (right) codes for the N20 model calibration. In the lower panels of Figure~\ref{fig:histo_N20} we show the cumulative distributions. The same plots are shown for W18 in Figure~\ref{fig:histo_W18}. The cumulative distributions coming from P-HOTB seem smoother given that we assigned a random uniformly distributed (within a range) value for each progenitor mass, as we explained above.

In this part of the analysis we did not include the nickel yield of SN~1992H from \citet{PP15a,PP15b} due to its value of $\sim0.28~M_{\odot}$, significantly higher than all the other SN~II in our sample. However, it follows the correlation shown in Figure~\ref{fig:mni} and is consistent with the range of values found in previous studies \citep{hamuy03,spiro14}. It is worth noting that \citet{schm94} and \citet{cloc96}  have calculated the distance to the host galaxy of SN~1992H, GC~5377, obtaining values of $\sim 30$ Mpc and $\sim 20$ Mpc (compared to $\sim 40$ Mpc with our method), respectively. These values give lower $\mni$ than ours for the same reddening, $\sim0.15~M_{\odot}$ and $\sim 0.06~M_{\odot}$, respectively. Nonetheless, the method used in this work is self-consistent.

\begin{figure*}[]
\epsscale{1.1}
\plotone{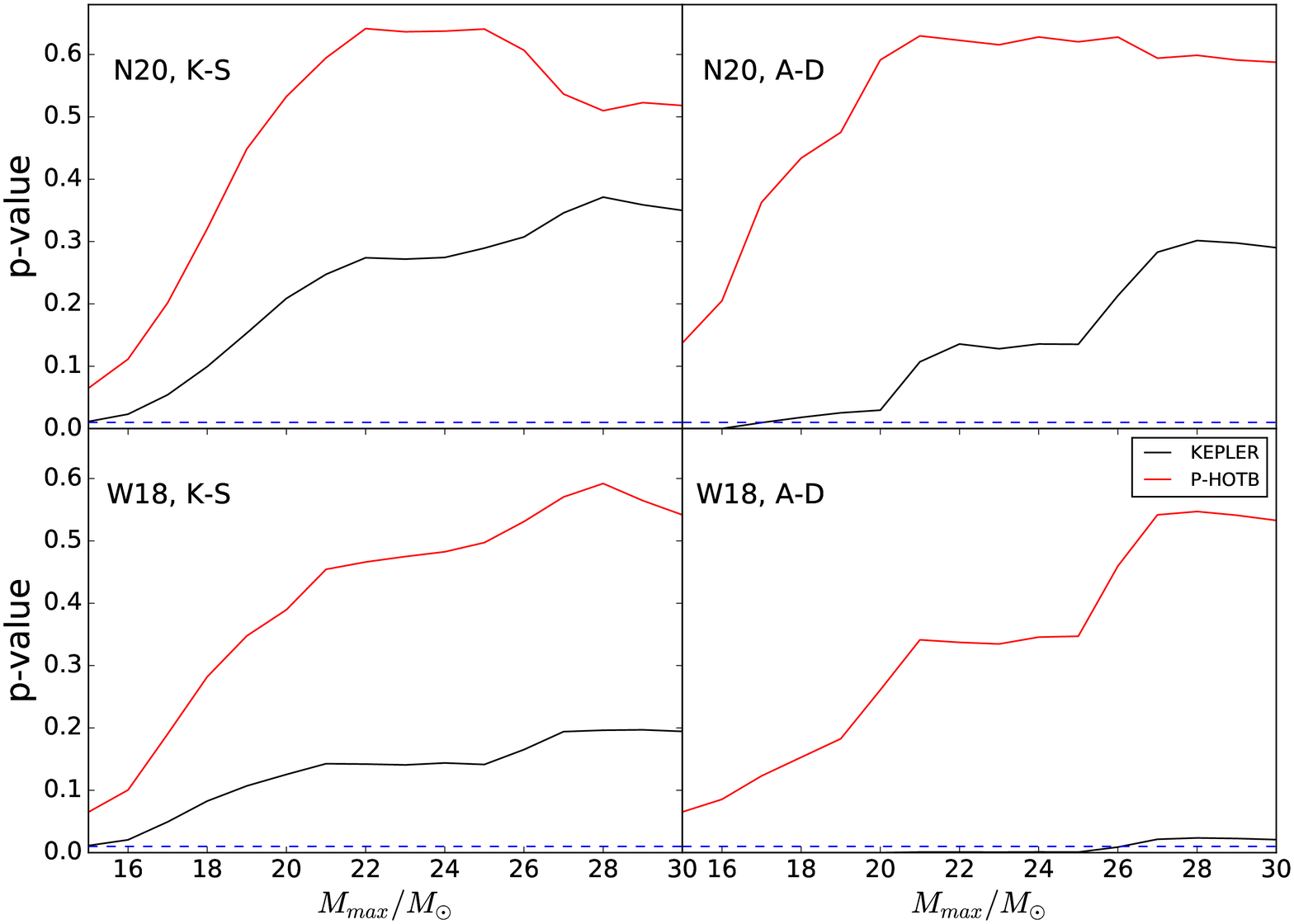}
\caption{p-values of the two-sample K-S tests (left panels) and A-D tests (right panels) between our joint sample and the different theoretical distributions coming from IMFs with different $M_{max}$. The N20 calibration is shown in the upper panels and the W18 calibration in the lower panels. p-values from KEPLER and P-HOTB distributions are shown in a solid black line and red line, respectively. The dashed blue lines represent the significance level of the tests, $p_{\rm threshold} = 0.01$. We can clearly see that most of the p-values from the P-HOTB distributions are higher than the ones from the KEPLER distributions. Only for the W18 calibration with KEPLER the p-values given by the A-D test are below $p_{\rm threshold}$}
\label{fig:p-v}
\end{figure*}

\section{Discussion and Conclusions}
\label{sec:conc}

We identify at least two different populations in the theoretical $\mni$ explosion models distributions for the two hydrodynamic codes presented in \citet{suk16}. We clearly see a small gap around $\mni \sim 0.06~M_{\odot}$ for the different models (see Figure~\ref{fig:histo_N20} and \ref{fig:histo_W18}). This is caused by the change in progenitor structure around $M \sim 12~M_{\odot}$ (see Figure 5 of \citealt{suk16}). Some parameters for the Crab-like progenitors are interpolated between the Crab model and the SN~1987A-like progenitors \citep[see Section 3.1.3 in][]{suk16}.


The $\mni$ distributions obtained from both codes and model calibrations (N20 and W18) seem to follow the trend of the observed distribution at different $\mni$ ranges. We also notice that the theoretical $\mni$ values have a maximum yield of $\mni \sim 0.08-0.12~M_{\odot}$, but the observations show some luminous SN~II that produce $\mni \gtrsim 0.28~M_{\odot}$  (e.g., SN~1992H in this study; see also the sample of \citealt{hamuy03}), much lower than typically high $\mni$ yields of $0.6-0.7~M_{\odot}$ of SNe type Ib and Ic \citep[e.g.,][]{Drout11,Prentice16}. Further theoretical work is needed to be able to explain the highest $\mni$ values seen in observations of normal SN~II. 

In addition, we compared the cumulative distributions of $\mni$ obtained from the P-HOTB and KEPLER codes, for both model calibrations (N20 and W18), with the observations. We performed two-sample Kolmogorov-Smirnov tests \citep[hereafter K-S;][]{Chakra67} and two-sample Anderson-Darling tests \citep[hereafter A-D;][]{Stephens74}. The K-S test lets us compare two samples and tells us if they could have been drawn from the same distribution, where a p-value is used to quantify this. The A-D test is a modified version of the K-S test that gives more weight to the tails of the distributions than the K-S test does. The p-value gives us the probability of obtaining a result equal to or more extreme than what we observe here. If this value is below a certain threshold, called the significance level ($p_{\rm threshold}$), we can discard our null hypothesis that both distributions, the observed and the theoretical one, come from the same distribution. We have set the value of $p_{\rm threshold}$ to $0.01$ \citep{WL16}.

For the analysis we did several two-sample K-S and A-D tests for distributions coming from IMFs with different upper-mass thresholds, $M_{\rm max}$, ranging from $15$ to $30~M_{\odot}$ (most of the progenitors above $30~M_{\odot}$ fail to explode or explode after losing their hydrogen envelopes, i.e, not as SN~II). These results are shown in Figure~\ref{fig:p-v}. For most of the cases we see that the p-values are well above $p_{\rm threshold}$ and we cannot exclude that observations and theory are drawn from the same distribution for nearly all choices of $M_{\rm max}$. The only exception is when we use the A-D test with the W18 calibration and KEPLER. This could be caused by the absence of high $\mni$ values for this calibration and the higher weight that the A-D test puts on the tails of the distribution. However, we think that this is not enough evidence to reject KEPLER model results. 

Taking these results at face value and given the conclusions from \citet{Smartt15}, that $M_{\rm max} \approx 18~\msun$, we might speculate that predictions of $\mni$ from neutrino mechanism are compatible with currently existing observations of SN~II. Nonetheless, we are not taking into account the uncertainties in the observed $\mni$ and biases due to sample completeness. A larger and more complete sample is also needed to have better statistics and to obtain more robust results, where ongoing (e.g., ASAS-SN) and future surveys (e.g., ZTF, LSST) will play a key role. It is possible that better scaling relations than \citet{litvinova85} and \citet{popov93}, that include a more realistic approach, will correlations between quantities such as $E_{\rm exp}$ and $M_{\rm env}$. We also need to take into account that 1D models take into account multi-dimensional fluid instabilities and mixing at best only approximately \citep[e.g.,][]{janka12,burrows13,takiwaki14} and a larger sample of progenitors and calibrations need to be tested since the ones used in this work may not be fully representative of the true nature of core-collapse explosions.

In this work we have doubled the sample of normal SN~II from \citet{PP15a,PP15b} and used the same code to fit multicolor light curves and expansion velocity curves. The slopes and intrinsic widths of the correlations between $\mni$, plateau luminosity at 50 days after explosion, and $\eexp$ are compatible with \citet{PP15b}. 

We studied the completeness of our joint sample and found that the sample has a slight deficit on the low luminosity end. More low luminosity supernovae with lower $\mni$ would perhaps improve the match in the lowest bin of $\mni$ distributions, so a larger sample is needed to increase the statistics.

We convolved a Salpeter IMF, with $M_{\rm min} = 9~M_{\odot}$ and $M_{\rm max}$ ranging from $15$ to $30~M_{\odot}$, with the progenitor masses from \citet{suk16} to retrieve a set of theoretical $\mni$ distributions from the N20, W18 and Z9.6 model calibrations presented in their work and for two different hydrodynamic codes, Prometheus-Hot Bubble (P-HOTB) and KEPLER. We compared these distributions with our joint sample through two-sample Kolmogorov-Smirnov test and two-sample Anderson-Darling test obtaining some slightly significant differences between both codes. The p-values calculated suggest that KEPLER and Prometheus Hot Bubble match the observational (joint) sample in general for different $M_{\rm max}$ and progenitor calibrations. 


Our work is one of the steps necessary to verify the explosion mechanism of CCSNe and this could help to understand the emerging theoretical pattern that success and failure of core collapse in massive stars depends sensitively on initial conditions and is not monotonic in the initial mass. The theoretical work on 1D parameterized models has provided a suite of observationally-testable predictions. Further progress in this area will likely require closer interaction of the parameterized models with multi-dimensional simulations.

\section*{Acknowledgements}
We thank M.~Hamuy and C.~Guti\'errez for providing the optical spectra for several of the SNe. We thank Tuguldur Sukhbold, Thomas Janka and Todd Thompson for discussions and detailed comments on the manuscript. We also thank the referee for detailed comments. TM, JLP, and AC were supported in part by the Ministry of Economy, Development, and Tourism's Millennium Science Initiative through grant IC120009, awarded to the Millennium Institute of Astrophysics, MAS. Support for JLP was also provided by FONDECYT through the grant 1151445. Support for OP was provided in part by NASA through Hubble Fellowship grant HST-HF-51327.01-A awarded by the Space Telescope Science Institute, which is operated by the Association of Universities for Research in Astronomy, Inc., for NASA, under contract NAS 5-26555. TM and AC were also supported by CONICYT through grant Basal CATA PFB 06/09.

\begin{appendix} 
\label{Appdx:1}
\begin{figure*}
\epsscale{1.17}
\plottwo{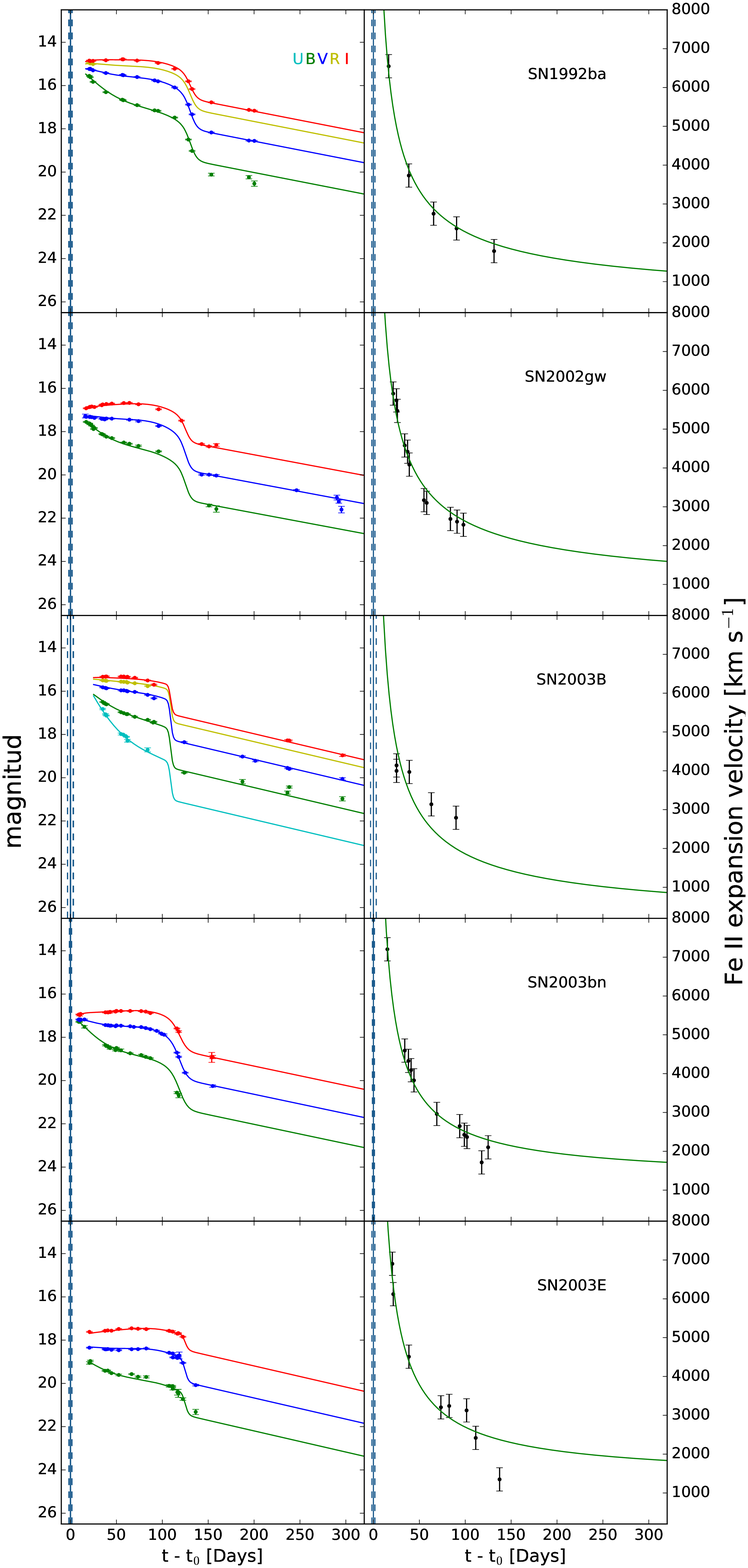}{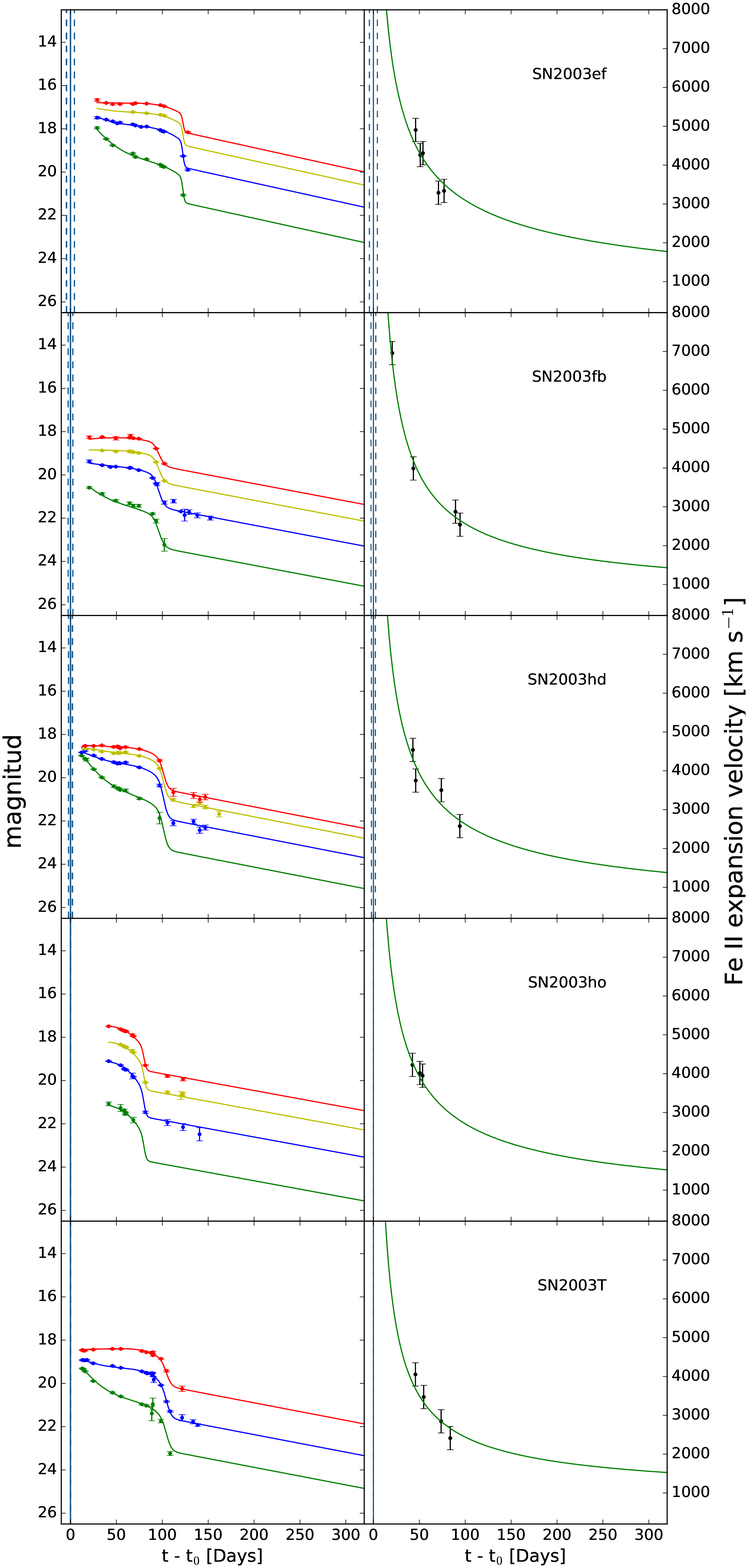}
\caption{Multiband light curves and photospheric expansion velocity curves along with their best-fit models for 10 out of the 16 SN~II (remaining objects are shown in Figs.~\ref{fig:curves_plot2} and \ref{fig:curves_plot3}). The left panels of both columns show the light curves in the optical bands while the right panels show the expansion velocity curves for each SN. The vertical solid blue lines represent the explosion times $\texp$ derived from the fits with their uncertainties (vertical dashed blue lines). Each SN has the same vertical and horizontal axes ranges.}
\label{fig:curves_plot}
\end{figure*}

\begin{figure*}
\epsscale{0.9}
\plotone{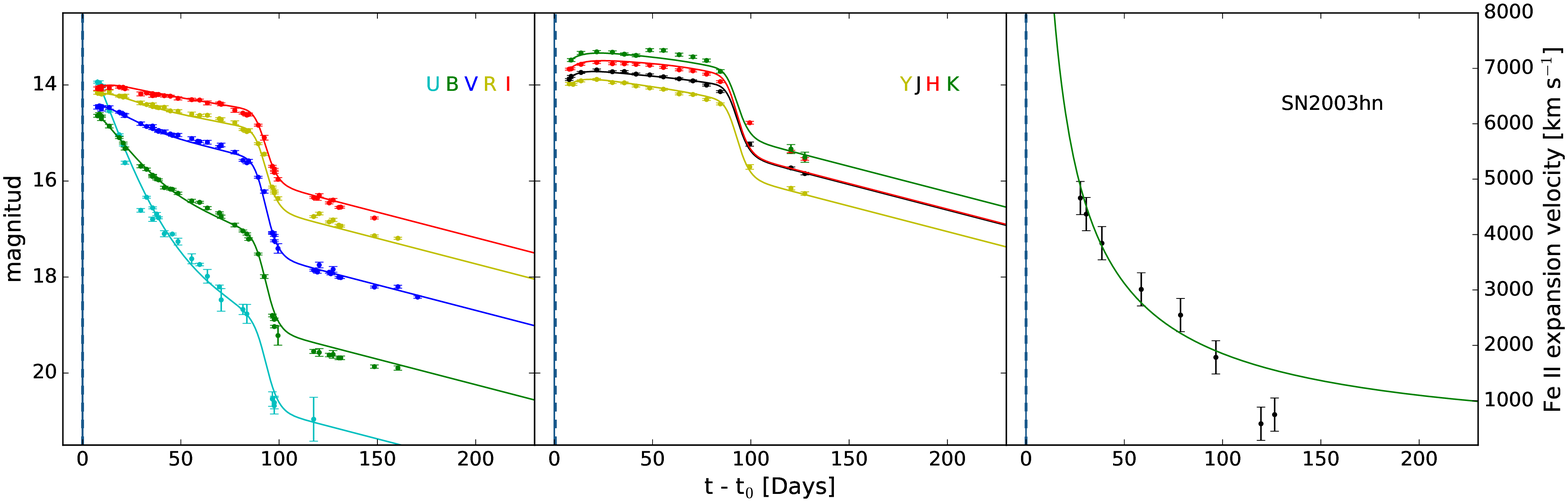}
\plotone{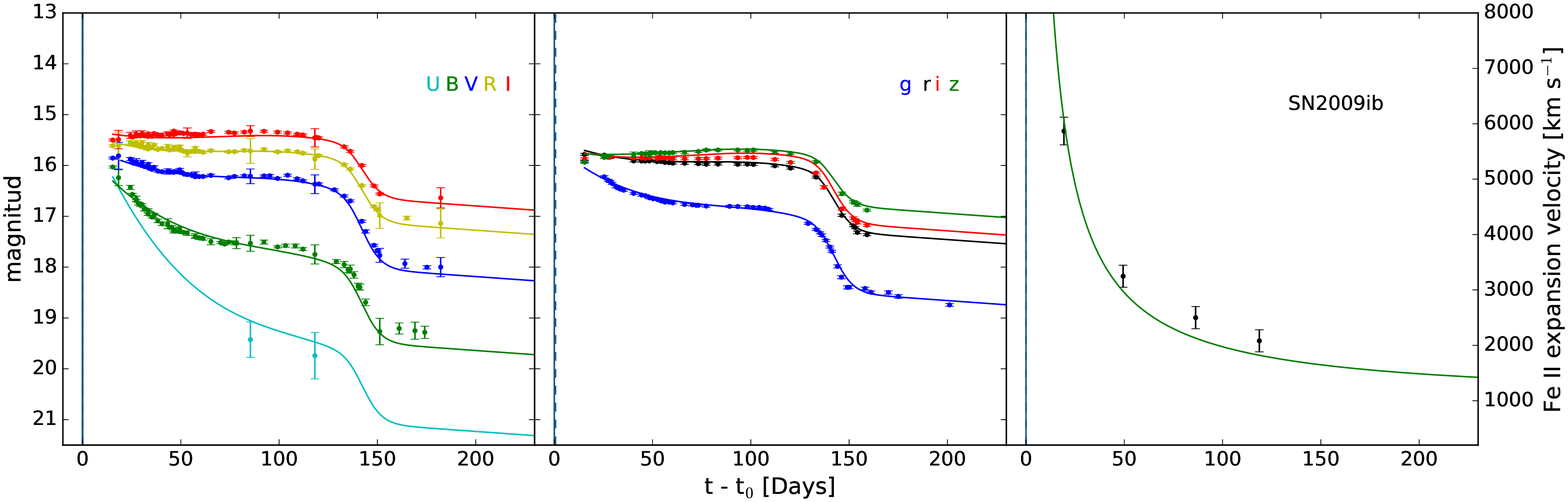}
\epsscale{1.15}
\plotone{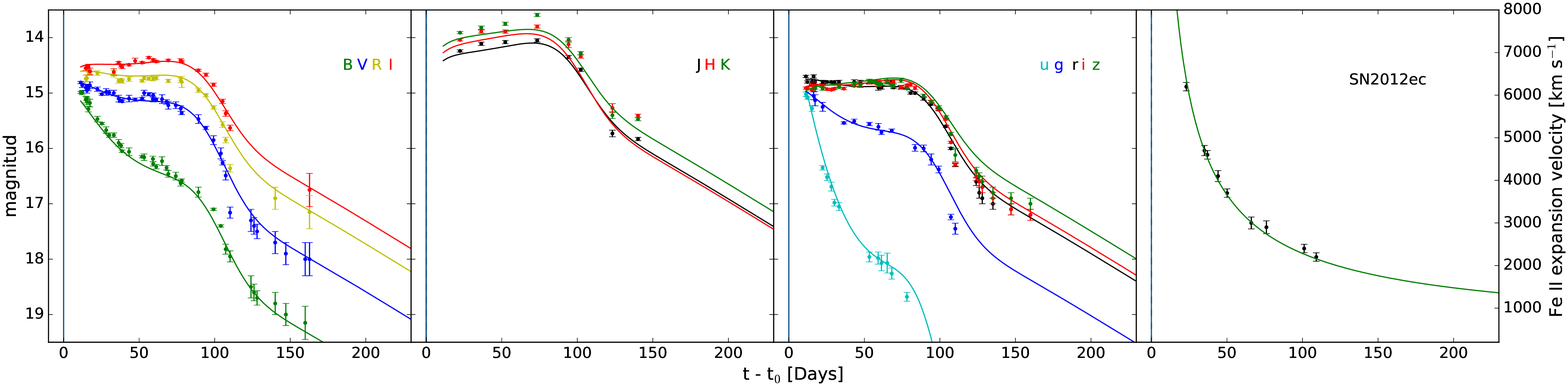}
\caption{Same as in Fig.~\ref{fig:curves_plot} but for SN~2003hn, SN~2009ib, and SN~2012ec.}
\label{fig:curves_plot2}
\end{figure*}

\begin{figure*}
\epsscale{0.55}
\plotone{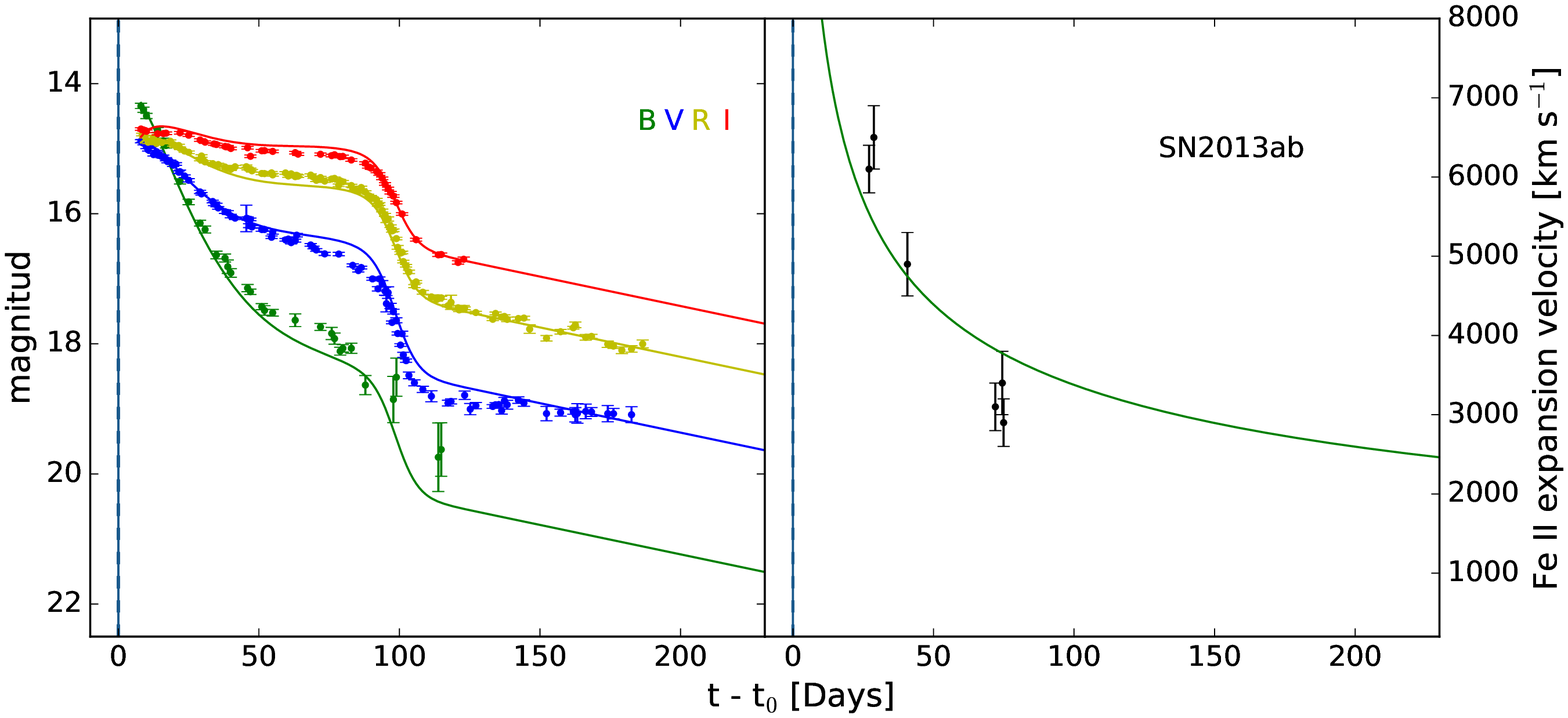}
\epsscale{1.15}
\plotone{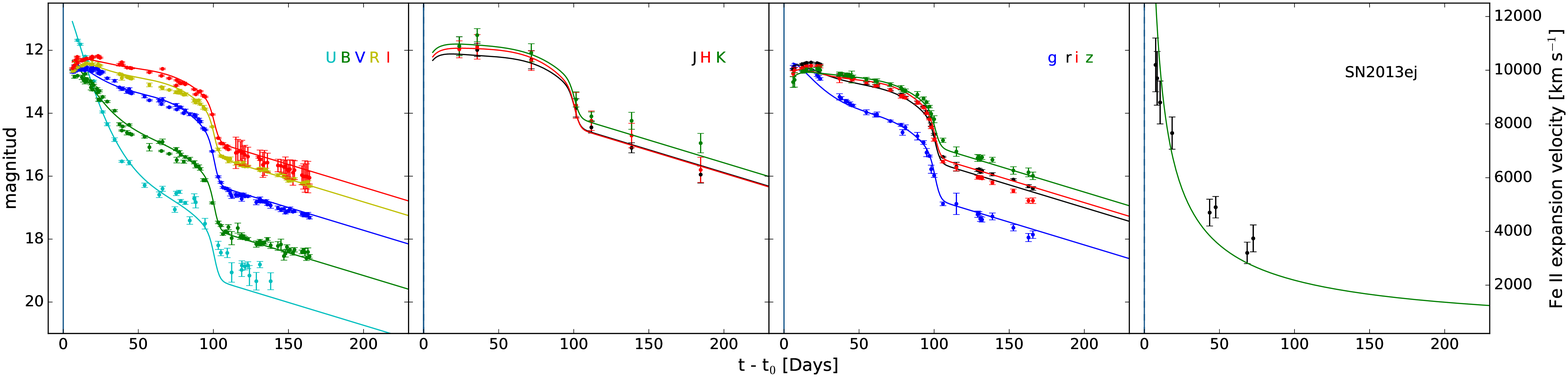}
\plotone{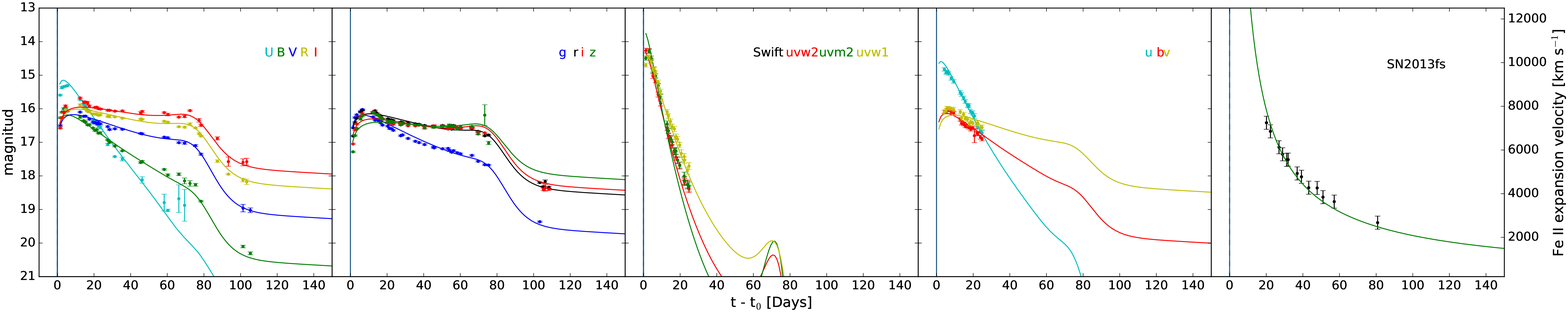}
\caption{Same as in Fig.~\ref{fig:curves_plot} but for SN~2013ab, SN~2013ej, and SN~2013fs. Note that the vertical axis differ for each object.}
\label{fig:curves_plot3}
\end{figure*}

\begin{figure*}
\epsscale{1.15}
\plotone{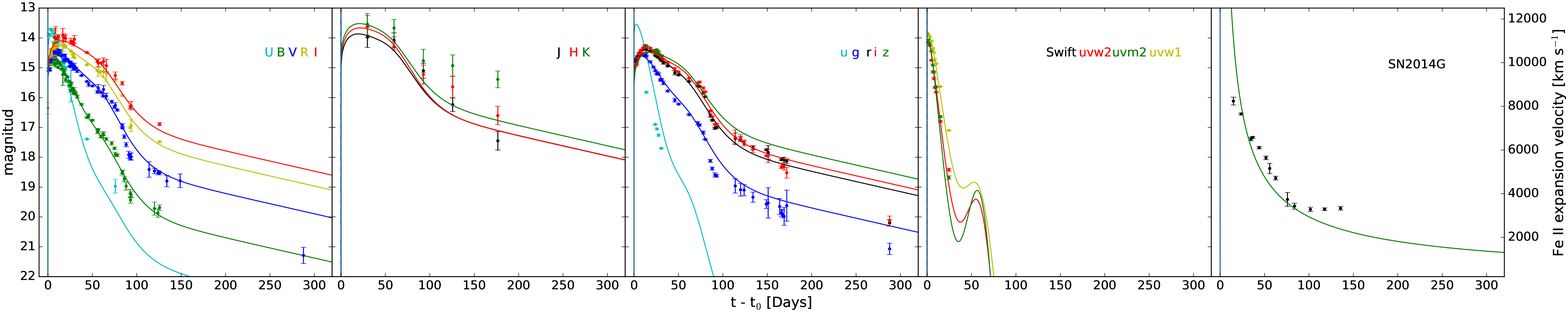}
\plotone{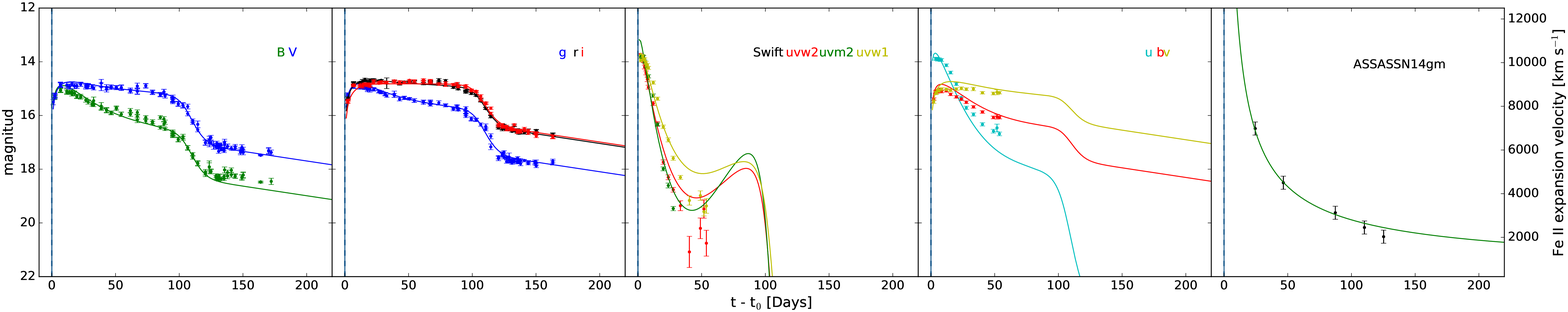}
\plotone{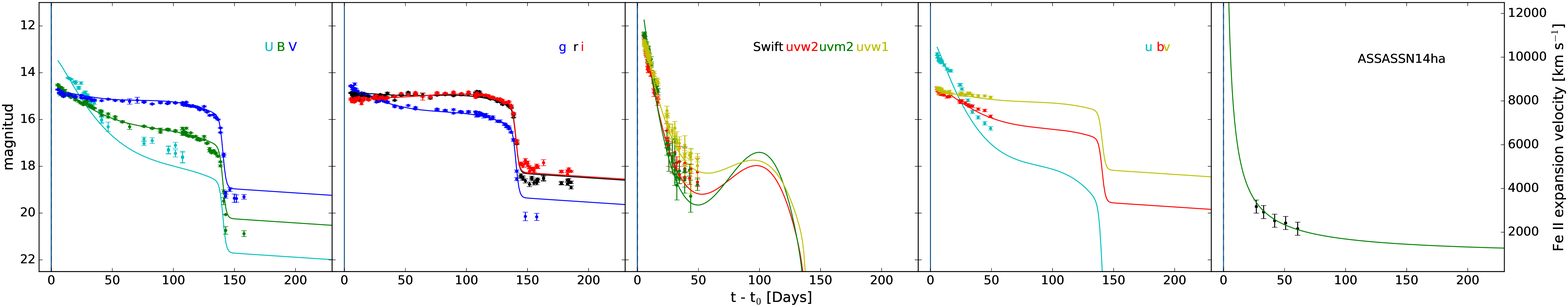}
\caption{Same as in Fig.~\ref{fig:curves_plot} but for SN~2014G, ASASSN-14gm, and ASASSN-14ha. Note that the vertical axis differ for each object.}
\label{fig:curves_plot4}
\end{figure*}

\end{appendix}

\end{document}